\documentclass[prd,showpacs,floatfix,amsmath,amssymb,nofootinbib,twocolumn]{revtex4-1}
\usepackage{graphicx,color,dcolumn,booktabs,bm}
\usepackage{longtable,comment,braket}
\usepackage{times}
\usepackage{overpic}
\usepackage{amssymb,tipa}
\usepackage{indentfirst}
\usepackage{feynmf}   
\usepackage{slashed}  
\usepackage{cases}
\usepackage{psfrag}
\usepackage{subfigure}
\usepackage{color}
\usepackage{multirow}
\usepackage{lineno}

\usepackage[colorlinks, citecolor=blue,anchorcolor=red,menucolor=red,linkcolor=blue,filecolor=red,runcolor=red,urlcolor=blue,frenchlinks=red]{hyperref}

\usepackage{ulem}
\def\be{\begin{equation}}
\def\ee{\end{equation}}

\begin{document}

\title{High-precise determination of critical exponents in holographic QCD}
\author{Fei Sun$^{a,d}$}\email{sunfei@ctgu.edu.cn}
\author{Xun Chen$^{b}$}\email{chenxun@usc.edu.cn}
\author{Shuang Li$^{a,d}$}\email{lish@ctgu.edu.cn}
\author{Akira Watanabe$^{c}$}\email{watanabe@usc.edu.cn}

\affiliation{$^a$College of Mathematics and Physics, China  Three Gorges University, Yichang, 443002, China \\
$^b$School of Nuclear Science and Technology, University of South China, Hengyang 421001, China
\\
$^c$School of Mathematics and Physics, University of South China, Hengyang 421001, China
\\
$^d$Center for Astronomy and Space Sciences, China Three Gorges University, Yichang 443002, China}

\begin{abstract}

The precise determination of critical exponents is crucial for understanding the properties of strongly interacting matter under extreme conditions. These exponents are fundamentally linked to the system's behavior near the critical end point (CEP), making precise localization of the CEP essential. However, precisely identifying the CEP within the framework of AdS/CFT correspondence presents considerable challenges. In this study, we explore critical phenomena and critical exponents within a holographic QCD model. We achieve high-precision calculations of the CEP's position and dynamically analyze the behavior of the critical exponent as it approaches the CEP using limiting methods. Our results indicate that linear fitting is only appropriate in regions very close to the CEP. Furthermore, we find that although the values of the critical exponents vary when approaching the CEP from different directions, they ultimately converge to a single fixed value, revealing a universal behavior among these exponents. Our research underscores the importance of precisely determining the CEP, selecting the fitting region, and considering the direction of the approach when predicting critical exponents. These findings establish a vital benchmark for identifying critical phenomena and their associated exponents.
\end{abstract}



\maketitle

\section{Introduction\label{sec1}}

In the field of fundamental particle physics, the study of the phase structure of quantum chromodynamics (QCD) is of paramount importance. QCD, the theory describing strong interactions, provides a framework for exploring phase transitions of matter under extreme conditions of temperature and density, particularly, the transition from hadronic matter to quark-gluon plasma (QGP). This research carries profound implications. On one hand, it enhances our understanding of the early evolution of the universe and the formation of elements, shedding light on the non-perturbative effects of strong interactions and elucidating the complex dynamics among particles. On the other hand, insights gained from studying the QCD phase structure offer valuable clues for the exploration of cosmology, gravitational waves, and new physics \cite{Espinosa:2010hh,Boeckel:2011yj,Caprini:2015zlo,Liu:2021svg,Shao:2022oqw,Pasechnik:2023hwv,Shao:2024dxt}. The study of QCD phase structure focuses on transitions between various states of matter under strong interactions, with particular emphasis on the hadronic-to-QGP transition. Within this context, the CEP is a central feature of the QCD phase diagram, marking the terminus of first-order phase transitions and the onset of continuous phase transitions, where some physical quantities, such as electrical conductivity, specific heat, etc., undergo drastic changes in response to variations in temperature and chemical potential.  Therefore, the critical behavior near the CEP is of particular interest, and the critical exponents quantify the drastic changes in different physical quantities near the CEP. Understanding the nature of these exponents and their interrelationships is crucial for studying the QCD phase transition and its physical characteristics.

Precisely determining the location and properties of the CEP is crucial for a comprehensive characterization of the QCD phase structure and for clarifying the nature of phase transitions. Currently, heavy-ion collision experiments at facilities such as the Relativistic Heavy Ion Collider (RHIC) and the Large Hadron Collider (LHC) serve as primary tools for empirically testing theoretical predictions. These experiments not only provide critical insights into key physical parameters related to the CEP, such as temperature and chemical potential, thereby validating theoretical models of QCD phase behavior but also deepen the scientific community's understanding of QCD phase structure, laying a robust foundation for advancements in particle physics.

Lattice QCD simulations have yielded extensive information on the QCD phase structure \cite{Aoki:2006we,Borsanyi:2013bia,HotQCD:2014kol,HotQCD:2018pds,Borsanyi:2020fev,Braguta:2021jgn,Braguta:2022str,Yang:2023vsw,Braguta:2024zpi,Borsanyi:2025dyp}. However, lattice calculations continue to face challenges, such as the sign problem at finite baryon chemical potential, despite notable progress \cite{Allton:2002zi,Allton:2005gk,Bazavov:2017dus,Borsanyi:2021sxv}. On the theoretical front, a variety of effective models and continuous field theory methods have been developed, including the Nambu--Jona-Lasinio (NJL) model, the Polyakov-Nambu--Jona-Lasinio (PNJL) model, the quark-meson model, and approaches based on Dyson-Schwinger equations (DSE) and functional renormalization group (FRG) techniques \cite{Hatsuda1994,Schwarz:1999dj,Zhuang:2000ub,Barducci:2004tt,Fukushima2004,Ratti2009,Brauner2010,Sasaki2011,Mao2013,Jiang:2016wvv,Sun:2020bbn,Sun:2023yux,Qiu:2023ezo,Bao:2024glw,Hua:2024bwn,Klevansky:1992qe,Buballa:2003qv,Kohyama:2015hix,Roberts:1994dr,Alkofer:2000wg,Cloet:2013jya,Schwarz:1999dj,Zhuang:2000ub,Chen:2014ufa,Chen:2015dra,Fan:2016ovc,Fan:2017mrk,Fu:2010ay,Bowman:2008kc,Schaefer:2004en,Schaefer:2007pw,Mao:2009aq,Schaefer:2011ex,Schaefer:2012gy,Qin:2010nq,Sun:2023kuu,Cao:2021rwx,Sun:2024anu,Luecker:2013oda,Fu:2016tey,Chen:2021iuo,Braun:2023qak}. These models are widely employed in analyzing the QCD phase diagram. Notably, the AdS/CFT correspondence \cite{Maldacena:1997re,Gubser:1998bc,Witten:1998qj} has significantly advanced the study of strongly coupled systems, establishing itself as a cutting-edge framework for uncovering the non-perturbative features of QCD and gaining considerable traction within the community. The development of the QCD phase diagram using a holographic approach was pioneered in Refs.  \cite{DeWolfe:2010he,DeWolfe:2011ts}, in which the Einstein-Maxwell-dilaton (EMD) theory was employed to replicate the characteristics of the QCD phase diagram \cite{Chelabi:2015cwn,Chelabi:2015gpc,Li:2016smq,Chen:2018msc,Chen:2019rez,Choun:2019xyo,Mamani:2020pks,Chen:2024jet,Ahmed:2024rbj,Wang:2024szr,Li:2024lrh,Zhao:2023gur,Rougemont:2023gfz,Zhao:2022uxc,Yang:2020hun}.

Despite these advancements, theoretical calculations often entail uncertainties. For instance, Refs. \cite{DeWolfe:2010he,Chen:2018msc,Chen:2018vty,Fu:2024wkn,Cai:2024eqa} report results consistent with mean-field theory, albeit with minor discrepancies. Recently, Ref. \cite{Zhao:2023gur} revealed that the critical exponent closely aligns with that of the quantum 3D Ising model, deviating significantly from mean-field theory predictions. This may highlights the interdependence between the critical exponent's numerical values and the precise location of the CEP, underscoring the need for rigorous theoretical and experimental efforts to accurately determine these parameters. Such precision is crucial for advancing our understanding of QCD's non-perturbative properties and for predicting the thermodynamic behavior of matter under varying temperature and chemical potential conditions \cite{Braun:2023qak,Cai:2024eqa}.

However, the precise localization of the CEP remains a challenging and dynamic area of research, with divergent predictions from various models and a lack of consensus on unified results. Current methods for determining critical exponents, such as linear fitting typically assume a linear relationship between variables, which may not hold unless the system is very near the critical point, potentially leading to flawed conclusions. The values of critical exponents can vary depending on the chosen fitting regions, and only within specific regions infinitely close to the critical point can data align precisely along a straight line. Thus, optimizing this method is crucial for avoiding drawing inaccurate conclusions. In this study, we aim to determine the CEP's location precisely while employing limiting methods to dynamically investigate the properties of critical exponents as the system approaches the CEP.

The quest for precise CEP determination is further motivated by experimental initiatives like the Beam Energy Scan (BES) program, which probes the QCD phase transition by systematically varying the collision energy in heavy-ion collisions. This enables the examination of the system's response as it approaches the CEP from different directions in the phase diagram. Such studies are critical, as the approach to the CEP may significantly influence critical exponent values. From a theoretical perspective, there is a limited discourse on calculating critical exponents under the AdS/CFT framework while considering different approaches to the CEP. This study addresses this gap by thoroughly calculating critical exponents as the system approaches the CEP from different directions. It addresses a pivotal question: how does the choice of the direction approaching the CEP affect the thermodynamic quantities and, in turn, how does it influence the calculated values of the critical exponents?

This study emphasizes the importance of precise critical exponent calculations. Traditional methods, which often neglect the CEP's precise location and rely on linear fittings within specific intervals, may introduce significant errors. Our objective is to provide a more comprehensive and theoretically consistent framework for studying critical phenomena by precisely determining the CEP's location and dynamically investigating critical exponent characteristics as the system approaches the CEP from different directions using limiting methods.

The organization of this work is as follows. Section \ref{sec2} provides a brief overview of AdS/CFT. Section \ref{sec3} presents numerical results for critical exponents. Finally, Section \ref{sec4} summarizes and discusses the findings.\\
\\

\section{model\label{sec2}}

We begin by introducing the general action of the 5-dimensional Einstein-Maxwell-Dilaton (EMD) system \cite{He:2013qq,Yang:2014bqa,Yang:2015aia,Dudal:2017max,Dudal:2018ztm,Chen:2018vty,Chen:2020ath,Zhou:2020ssi,Chen:2019rez}:
\begin{equation}
\begin{aligned}
S_{E} & =\frac{1}{16 \pi G_5} \int d^5 x \sqrt{-g}\left[R-\frac{f(\phi)}{4}F^2-\frac{1}{2} \partial_\mu \phi \partial^\mu \phi-V(\phi)\right], \\
\end{aligned}
\label{SE}
\end{equation}
where \(G_5\) is the effective Newton constant in five dimensions, \(g\) denotes the determinant of the metric tensor, and \(R\) represents the Ricci scalar, which characterizes the strength of the gravitational field. The second term describes the coupling between the dilaton field \(\phi\) and the gauge field, with \(f(\phi)\) being the gauge kinetic function. Here, \(F^2 = F_{\mu\nu}F^{\mu\nu}\), where \(F_{\mu\nu} = \partial_{\mu}A_{\nu} - \partial_{\nu}A_{\mu}\) is the field strength tensor of the Maxwell gauge field \(A\). The last two terms represent the kinetic term and potential energy of the dilaton field, respectively. The explicit forms of the gauge kinetic function \(f(\phi)\) and the dilaton potential \(V(\phi)\) can be consistently determined from the equations of motion (EOM).

To study QCD at finite temperature and chemical potential, we adopt the following asymptotic AdS black hole metric ansatz:
\begin{equation}
d s^2=\frac{L^2 e^{2 A(z)}}{z^2}\left[-g(z) d t^2+\frac{d z^2}{g(z)}+d \vec{x}^2\right],
\end{equation}
where \(L\) is the radial scale parameter of the \(\rm AdS_5\) space, \(z\) is the fifth-dimensional holographic coordinate, and \(A(z)\) describes the deformation of spacetime in the fifth dimension, which adjusts the metric components in different coordinate directions. Similarly, \(g(z)\) is a function of \(z\) that appears in the temporal component of the metric and contributes to the geometric structure of spacetime. By imposing regular boundary conditions at the horizon \(z = z_h\) and the asymptotic AdS condition at the boundary \(z = 0\), we obtain:
\begin{equation}
A(0)=-\sqrt{\frac{1}{6}} \phi(0), \quad g(0)=1, \quad A_t(0)=\mu+\rho^{\prime} z^2+\cdots.
\end{equation}
Here, \(\mu\) represents the baryon chemical potential, and \(\rho^{\prime}\) is proportional to the baryon number density.

Using the variational method, we derive the EOM from the action in Eq.~(\ref{SE}). Solving these equations yields:
\begin{widetext}
\begin{equation}
g(z) = 1 - \frac{{{I_1}\left( {0,z} \right)}}{{{I_1}\left( {0,{z_h}} \right)}} + \frac{{2c{\mu ^2}{e^k}}}{{{I_1}\left( {0,{z_h}} \right){{\left( {1 - {e^{ - cz_h^2}}} \right)}^2}}}\left| {\begin{array}{*{20}{l}}
{{I_1}\left( {0,{z_h}} \right)}&{{I_2}\left( {0,{z_h}} \right)}\\
{{I_1}\left( {{z_h},z} \right)}&{{I_2}\left( {{z_h},z} \right)}
\end{array}} \right|.
\label{gz}
\end{equation}
For simplicity, we define the following integrals:
\begin{equation}
{I_1}\left( {{z_1},{z_2}} \right) = \int_{{z_1}}^{{z_2}} d x{x^3}{e^{ - 3A(x)}},{I_2}\left( {{z_1},{z_2}} \right) = \int_{{z_1}}^{{z_2}} d y{y^3}{e^{ - 3A(y)}}{e^{ - c{y^2}}}.
\end{equation}
\end{widetext}

To obtain an analytical solution, we assume specific forms for \(f(\phi)\) and \(A(z)\) with several parameters. For the metric, we adopt the ansatz \cite{Chen:2020ath}:
\begin{equation}
A(z)= d \ln(a z^2 + 1) + d \ln(b z^4 + 1),
\label{eq:az}
\end{equation}
and for the gauge kinetic function \(f(z)\), we take:
\begin{equation}
f(z)=e^{c z^2-A(z)+k}.
\label{eq:fz}
\end{equation}

\begin{widetext}
The six parameters \(a\), \(b\), \(c\), \(d\), \(k\), and \(G_5\) are determined by machine learning the lattice QCD results for the equation of state (EOS) and baryon number susceptibility at zero chemical potential. Machine learning has become increasingly important in high-energy physics research and can also be applied within the AdS/CFT framework \cite{Hashimoto:2018ftp,Akutagawa:2020yeo,Hashimoto:2018bnb,Yan:2020wcd,Hashimoto:2021ihd,Song:2020agw,Zhou:2023pti,Ahn:2024gjf,Gu:2024lrz,Chen:2024ckb,Chen:2024mmd,Zhu:2025gxo}. Through machine learning methods, Ref. \cite{Chen:2024ckb} implements a deep neural network for regression analysis using the TensorFlow framework, which combines a machine learning-based approach with the EMD framework to construct an analytical form of the holographic black hole metric.

From the definition of baryon number density in Refs. \cite{Critelli:2017oub,Zhang:2022uin}, we have:
\begin{equation}
\rho {\rm{ }} = \left| {\mathop {\lim }\limits_{z \to 0} \frac{{\partial {{\cal L}^{\cal E}}}}{{\partial \left( {{\partial _z}{A_t}} \right)}}} \right| =  - \frac{1}{{16\pi {G_5}}}\mathop {\lim }\limits_{z \to 0} \left( {\frac{{{{\rm{e}}^{A(z)}}}}{z}f(\phi )\frac{{\rm{d}}}{{{\rm{d}}z}}{A_t}(z)} \right) =  - \frac{1}{{8\pi {G_5}}}\left( {\frac{{c{e^k}\mu }}{{1 - {e^{ - cz_h^2}}}}} \right).
\end{equation}
The black hole temperature and entropy density can be derived from Eq.~(\ref{gz}) at the event horizon \(z = z_h\):
\begin{equation}
T = \frac{{z_h^3{e^{ - 3A\left( {{z_h}} \right)}}}}{{4\pi {I_1}\left( {0,{z_h}} \right)}}\left( {1 + \frac{{2c{\mu ^2}{e^k}\left( {{e^{ - cz_h^2}}{I_1}\left( {0,{z_h}} \right) - {I_2}\left( {0,{z_h}} \right)} \right)}}{{{{(1 - {e^{ - cz_h^2}})}^2}}}} \right),  ~~~~~~ s=\frac{e^{3 A\left(z_h\right)}}{4 G_5 z_h^3}. \end{equation}
We also obtain:
\begin{equation}
\mu  = \left( {1 - {e^{ - cz_h^2}}} \right)\sqrt {\frac{{4\pi T{I_1}\left( {0,{z_h}} \right)}}{{z_h^3{e^{ - 3A({z_h})}}}} - 1} \sqrt {\frac{1}{{2c{e^k}\left( {{e^{ - cz_h^2}}{I_1}\left( {0,{z_h}} \right) - {I_2}\left( {0,{z_h}} \right)} \right)}}}. 
\end{equation}
\end{widetext}

\section{NUMERICAL RESULTS AND DISCUSSION \label{sec3}}
 For numerical calculations, in 2-flavor case, we choose the parameters as follows: $a=0.067~\text{GeV}^{2}, b=0.023~\text{GeV}^{4}, c=-0.377~\text{GeV}^{2}, d=-0.382, k= 0, G_5=0.885 ~\text{GeV}^{3}$, $L=1~\text{GeV}^{-1}$.  For more details, see Refs. \cite{Chen:2024ckb,Chen:2024mmd}. In this case, when the values are rounded to seven significant figures,  the CEP is located at $T_{CEP}=0.1471271
 $ GeV and $\mu_{CEP}=0.4598318$ GeV.
 
\begin{figure}
    \centering
    \includegraphics[width=9cm]{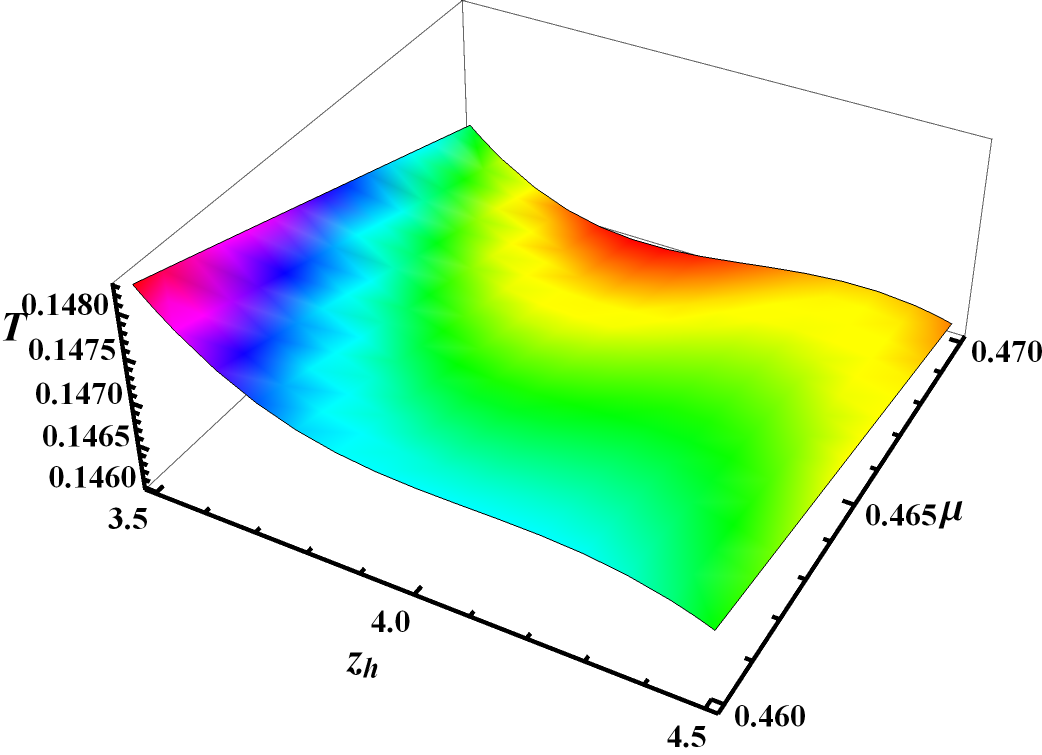}
    \caption{\label{Tmjuzh} The temperature $T$ depends on the horizon $z_h$ and the quark chemical potential $\mu$. The unit of $T$ or $\mu$ is $\text{GeV}$, the unit of $z_h$ is $\text{GeV}^{-1}$.}
    \label{Tmjuzh.pdf}
\end{figure}

\begin{figure}
\subfigure[]
{\includegraphics[width=0.45\textwidth]{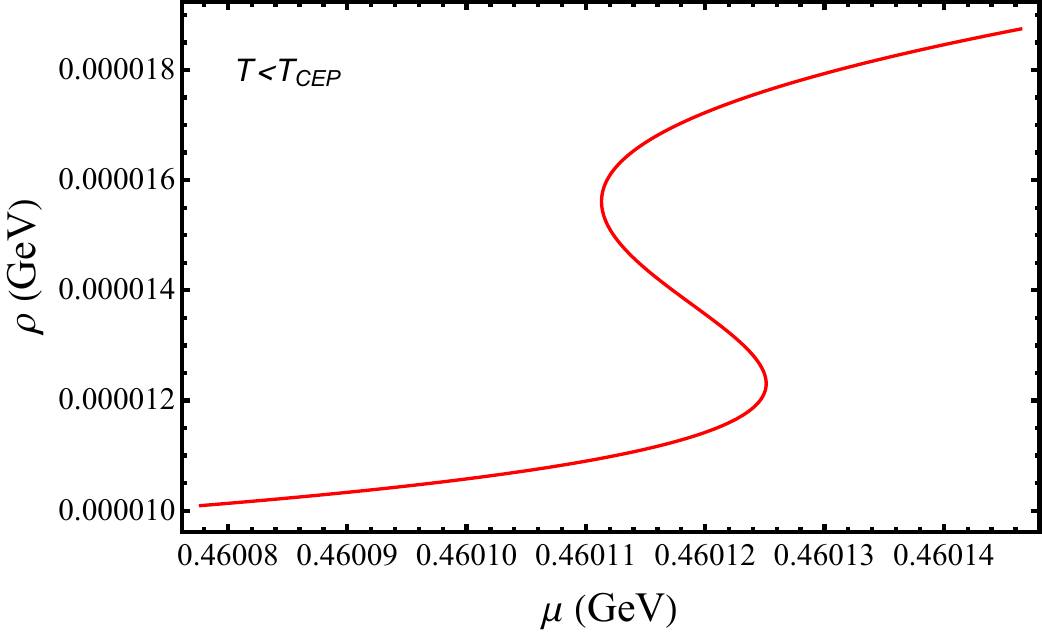}}
\subfigure[]
{\includegraphics[width=0.45\textwidth]{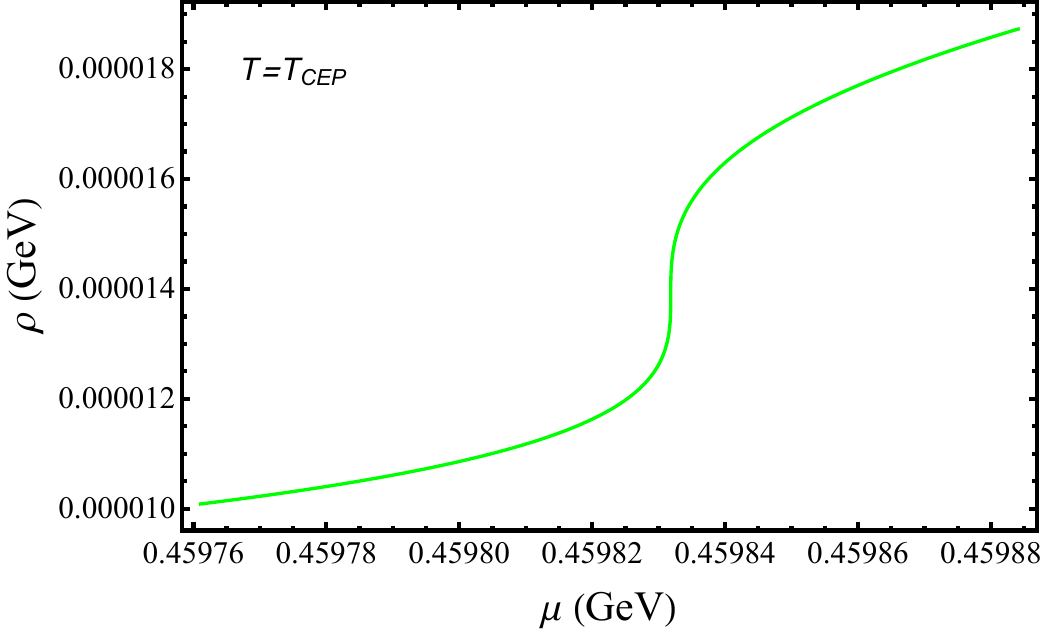}}
\subfigure[]
{\includegraphics[width=0.45\textwidth]{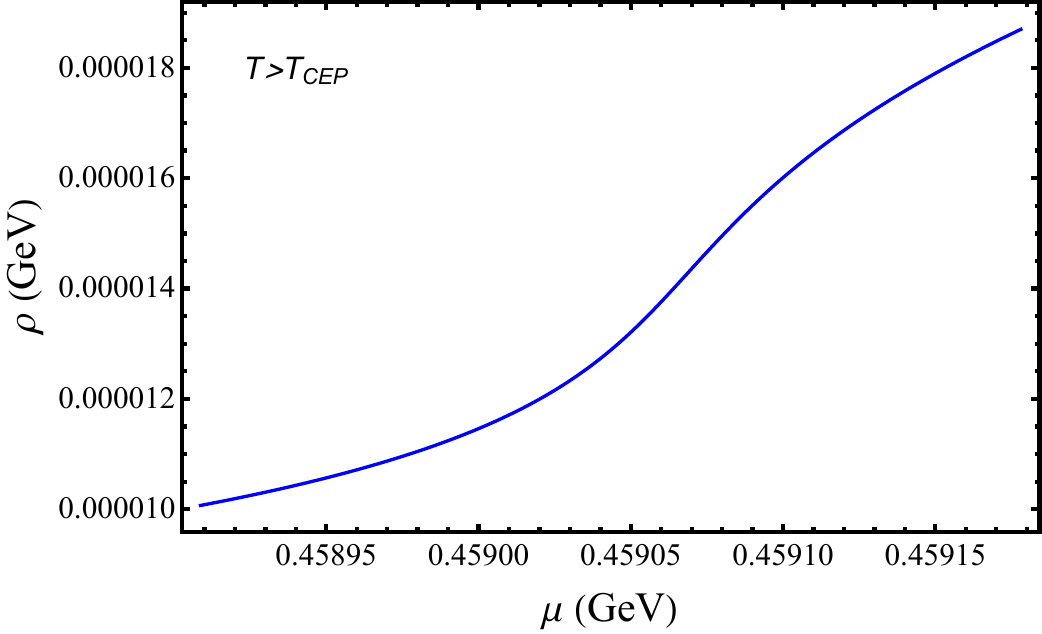}}
\caption[]{(Color online) The baryon number density $\rho$ as a function of the quark chemical potential $\mu$ for different temperatures around the CEP.}
\label{mjupho.pdf}
\end{figure}

Fig. \ref{Tmjuzh.pdf} illustrates the three-dimensional variation of temperature with respect to the fifth-dimensional holographic coordinate and the chemical potential. The behavior is notably complex. When the chemical potential is held constant, the temperature exhibits an upward trend in certain regions as the holographic coordinate increases. In other regions, the temperature may increase, decrease, or display more intricate fluctuations. This complexity likely arises from the influence of the holographic coordinate on the system's energy distribution, thereby affecting the temperature. When the black hole horizon \(z_{h}\) is fixed, the temperature decreases as the chemical potential increases. In the context of black holes, an increase in the chemical potential alters the effective potential experienced by the field configurations within the system, driving the black hole to stabilize at lower temperatures as it absorbs more charge from the thermodynamic state.

In Figure \ref{mjupho.pdf}, the baryon number density \(\rho\) is plotted as a function of the quark chemical potential \(\mu\) near the critical temperature \(T_{CEP }\). We consider three scenarios: \(T < T_{CEP }\), \(T = T_{CEP }\), and \(T > T_{CEP }\). For \(T < T_{CEP }\), the baryon number density as a function of the quark chemical potential exhibits multivalued behavior. This behavior indicates that a single value of \(\mu\) can correspond to two distinct values of \(\rho\): one associated with the hadronic phase and the other with the QGP phase. As the system approaches the CEP, competition between different phases emerges, indicating a first-order phase transition. This suggests that the system is undergoing a phase transition into a symmetry-broken phase, which is characterized by quark condensation and spontaneous symmetry breaking. At \(T = T_{CEP }\), the slope of the baryon number density exhibits significant changes, with the slope tending to infinity at \(\mu = \mu_{CEP }\). This marks the system's transition from one phase to another. For \(T > T_{CEP }\), the discontinuity associated with the first-order phase transition vanishes, and the system enters the crossover region, where the baryon number density increases smoothly with the quark chemical potential. This indicates a degree of symmetry restoration, transitioning from baryonic matter (ordered phase) to the QGP (disordered phase).

The CEP is a unique feature of the QCD phase diagram, and the physical behaviors in its vicinity are essential for understanding the properties of QCD matter. Critical exponents serve as key indicators to characterize the properties of the CEP. By studying these exponents, we can gain profound insights into how the degrees of freedom of quarks and gluons evolve near the CEP and how their interactions influence the macroscopic properties of the system. This, in turn, elucidates the microscopic physical mechanisms underlying QCD. First, we focus on the critical exponent \(\alpha\), which describes the behavior of the specific heat capacity \(C_\rho\) at constant baryon density \(\rho\) as the system approaches the critical point along the first-order phase transition line. The exponent \(\alpha\) characterizes the power-law behavior of the specific heat near the critical temperature \(T_c\):
\begin{eqnarray}
C_\rho \sim |T - T_c|^{-\alpha},
\end{eqnarray}
where the specific heat capacity \(C_\rho\) is defined as:
\begin{eqnarray}
C_\rho \equiv T \left( \frac{\partial s}{\partial T} \right)_\rho.
\end{eqnarray}

\begin{figure}
{\includegraphics[width=0.45\textwidth]{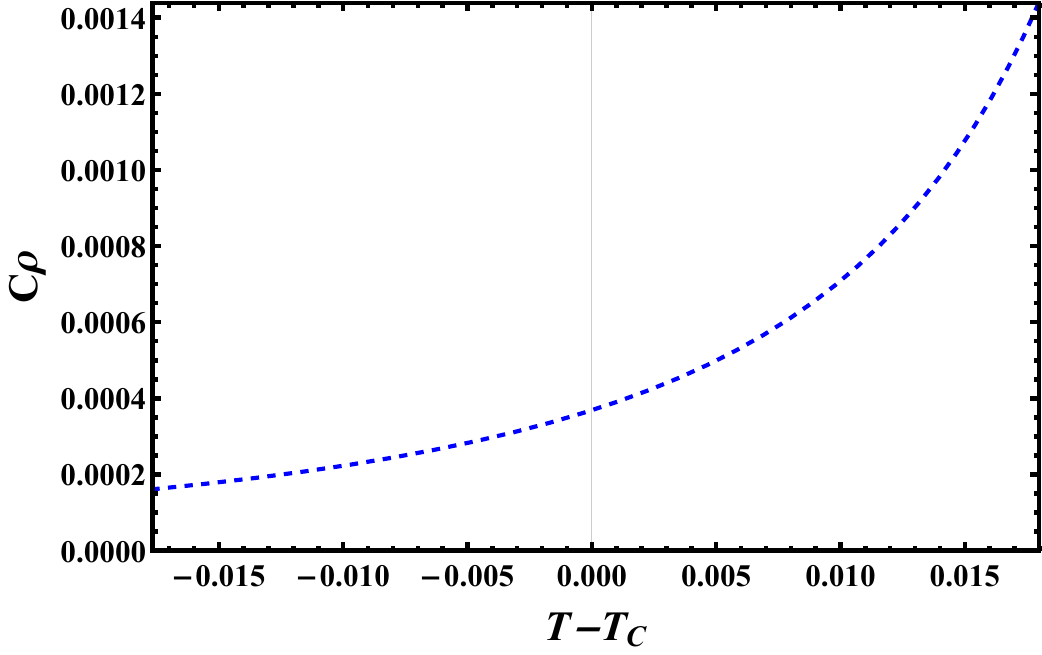}}
\caption[]{(Color online) The specific heat as a function of temperature
at fixed  baryon number density.}
\label{TCvpho.pdf}
\end{figure}

\begin{figure}
{\includegraphics[width=0.48\textwidth]{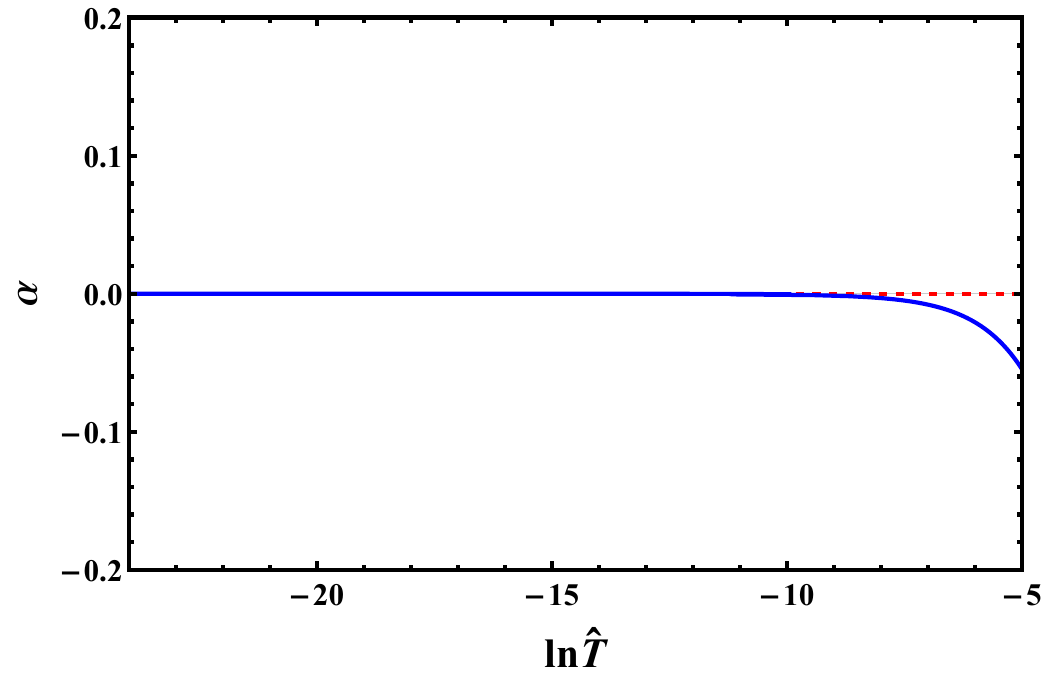}}
\caption[]{(Color online) The  critical exponent  $\alpha$ as a function of reduced temperature $\ln \hat{T}$ at fixed  baryon number density.}
\label{Talpharight.pdf}
\end{figure}

\begin{figure}
\subfigure[]
{\includegraphics[width=0.45\textwidth]{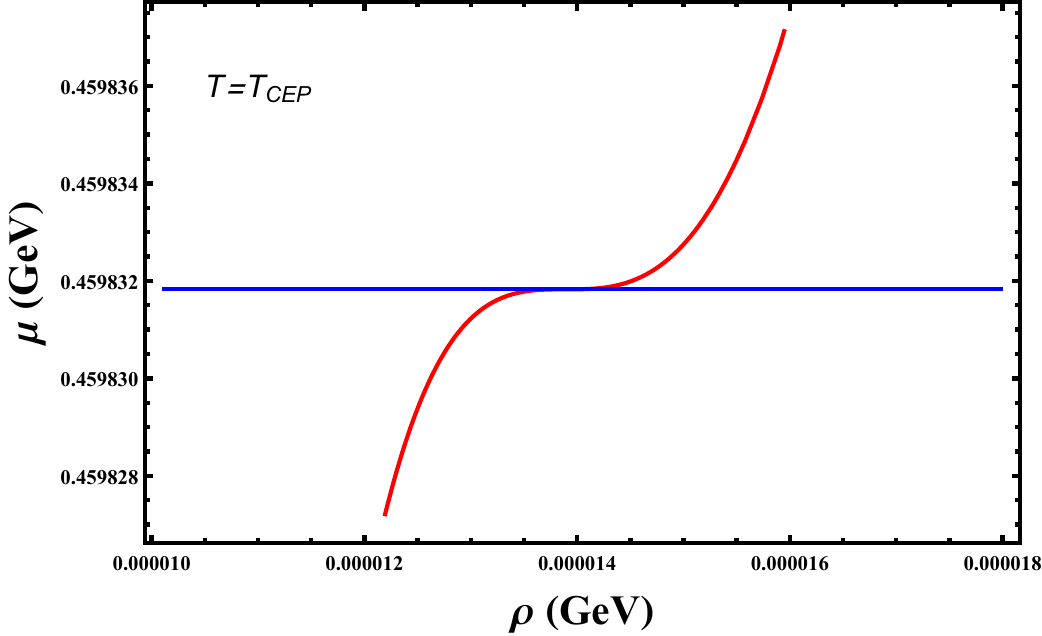}}
\subfigure[]
{\includegraphics[width=0.45\textwidth]{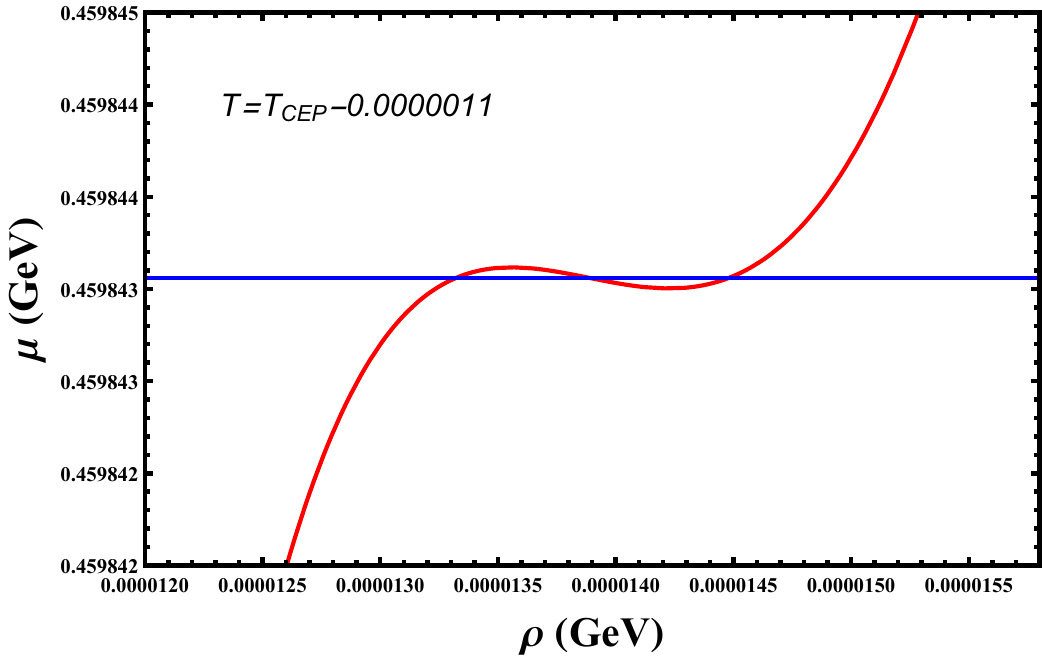}}
\subfigure[]
{\includegraphics[width=0.45\textwidth]{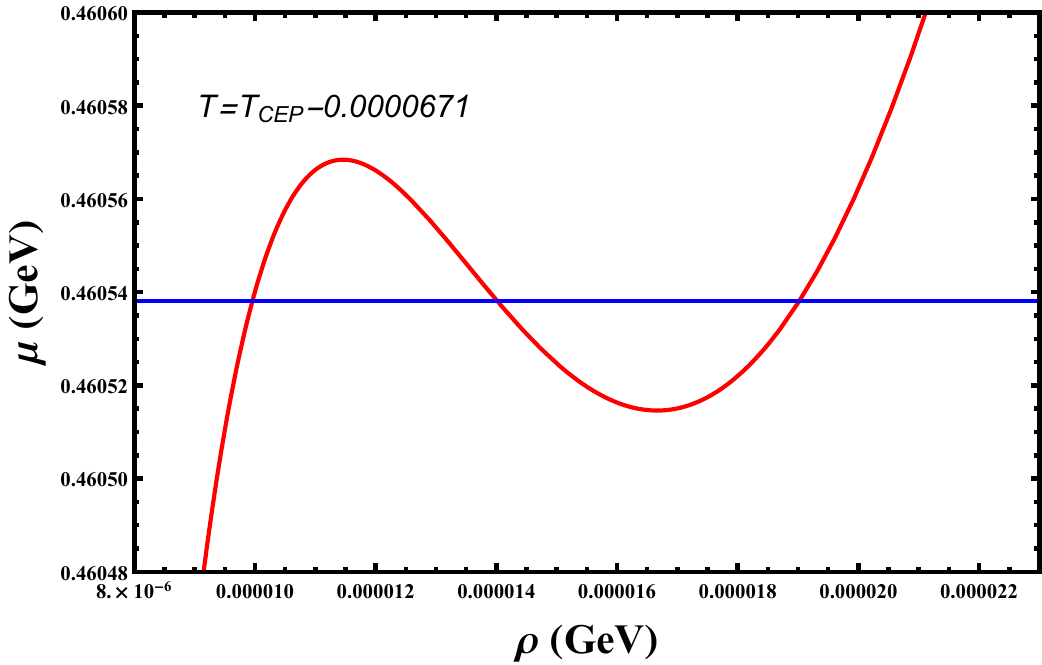}}
\caption[]{(Color online) The chemical potential as a function of baryon number density for different temperatures around $T_{CEP}$.}
\label{phomjuTcaround}
\end{figure}

In Fig. \ref{TCvpho.pdf}, we depict the specific heat as a function of temperature at a fixed baryon number density. The influence of approaching the CEP from different directions on the specific heat capacity \(C_{\rho}\) at constant baryon density \(\rho\) exhibits distinct behaviors. When approaching the CEP from the left (\(T < T_{CEP}\)), the system resides in the low-temperature phase. In this regime, \(C_{\rho}\) varies relatively slowly with temperature, and the corresponding critical exponent \(\alpha\) reflects the system's characteristics as it approaches the critical state in the low-temperature phase. Conversely, when approaching the CEP from the right (\(T > T_{CEP}\)), the system is in the high-temperature phase, where \(C_{\rho}\) changes more rapidly with temperature. When approaching the CEP from the left, experimental observations reveal gradual changes in the properties of hadronic matter, such as particle yields and momentum distributions. These changes are closely related to the value of \(\alpha\), which provides insights into the properties of hadronic matter in the low-temperature phase and its evolution toward the critical state. On the other hand, when approaching from the right, changes in the characteristics of high-temperature matter, such as QGP, can be observed. The value of \(\alpha\) in this regime aids in analyzing the physical processes associated with the transition from the high-temperature phase to the critical state, offering crucial information for understanding the QCD phase diagram.

Understanding the behavior of specific heat near the CEP is essential for studying phase transition phenomena. In Fig. \ref{Talpharight.pdf}, we plot the critical exponent  $\alpha$  which corresponds to the ratio \(\frac{\Delta \ln C_{\rho}}{\Delta \ln \hat{T}}\) as a function of the reduced temperature \(\ln \hat{T}\) at a fixed baryon number density. To illustrate the behavior of specific heat near the CEP, we employ logarithmic coordinates. Additionally, to ensure the horizontal axis is dimensionless, we introduce the reduced temperature \(\hat{T} = \frac{T - T_{CEP}}{T_{CEP}} \). It is clear to see that the value of critical exponent $\alpha$ is close to the result of the mean-field theory.  By analyzing the changes in the slope at different values of \(\hat{T}\), we gain deeper insights into the behavior of matter undergoing phase transitions.
Such a plot illustrates the dynamic changes in the slope with respect to \(\ln \hat{T}\) within a power-law relationship, enabling a more nuanced analysis of the relationships between variables. This is particularly relevant when the power-law relationship deviates from a simple linear pattern with a fixed slope, such as in cases where curvature or trends are present in the data.

To calculate the critical exponent \(\beta\) (discussed in detail later), we first introduce Maxwell's equal-area law. This law states that at the phase transition point, the free energies of the two phases are equal, thereby defining the phase transition line and the critical point. It is a fundamental tool in thermodynamics and phase transition studies, providing a practical method for analyzing first-order phase transitions and critical phenomena. However, its application requires careful attention to its range of validity and the precision of numerical computations. In practice, the implementation of the equal-area law necessitates highly accurate numerical calculations, as even minor errors can lead to incorrect determinations of phase transition conditions. Using Maxwell’s equal-area law, we determine the critical chemical potential \(\mu_c\). At the phase transition point, the free energies of the two phases must be equal, which implies that the areas bounded by \(\mu(\rho)\) and the line separating the two phases are also equal. In Fig. \ref{phomjuTcaround}, we present the chemical potential \(\mu\) as a function of baryon number density \(\rho\) for temperatures around \(T_{CEP}\). For computational convenience, the horizontal and vertical axes represent \(\rho\) and \(\mu\), respectively. Fig. \ref{phomjuTcaround} (a) shows the case at \(T = T_{CEP}\), and the blue line corresponds to the  \(\mu = \mu_{CEP}\). As illustrated in Figs. \ref{phomjuTcaround} (b) and \ref{phomjuTcaround} (c), the blue lines represent the quark chemical potentials at temperatures below \(T_{CEP}\), ensuring that the areas on either side of these curves are equal. To determine the first-order phase transition curve precisely, we meticulously calculate the areas of these regions. By numerically computing \(\mu_c\), we ensure that the areas of the two regions are equal. It is crucial to control the computational error to a very small value during these calculations.

\begin{figure}
{\includegraphics[width=0.48\textwidth]{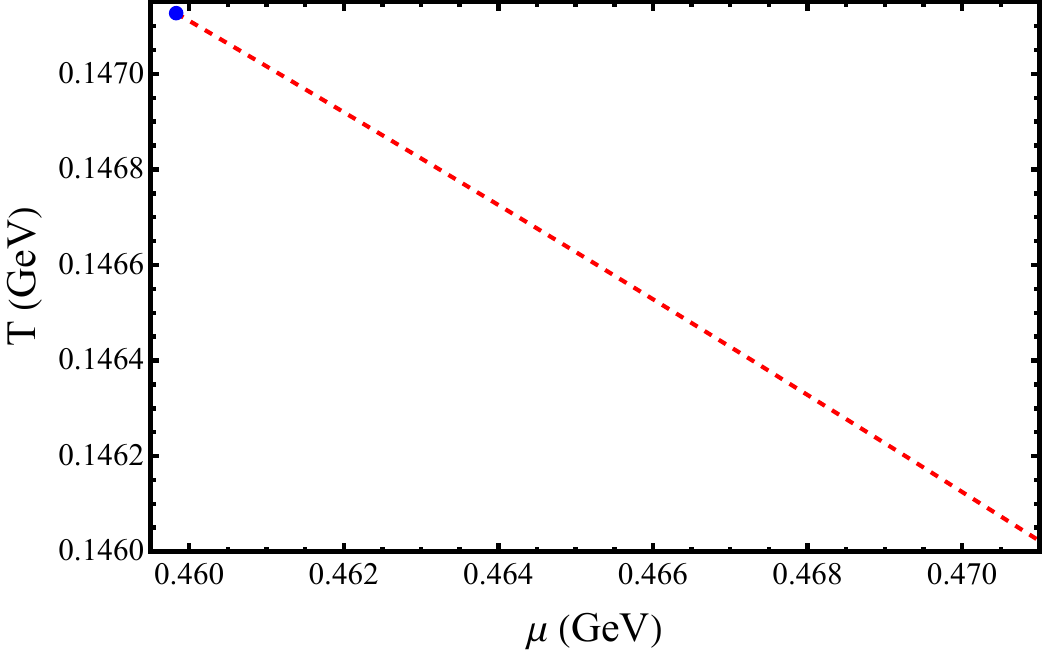}}
\caption[]{(Color online) The phase diagram around the CEP.}
\label{Tmju.pdf}
\end{figure}

In Fig. \ref{Tmju.pdf}, we present the phase diagram around the CEP using Maxwell's equal-area law. The blue dot here represents the CEP's position, and the red dashed line represents the first-order phase transition line. Accurately determining the first-order phase transition is of paramount importance in high-energy physics, as it enhances our understanding of the QCD phase diagram, including the properties of the QGP and the mechanisms of phase transitions between the QGP and hadronic matter. Additionally, precise knowledge of the first-order phase transition boundary is essential for locating the QCD critical point. Furthermore, the chemical freeze-out line may intersect or lie close to the first-order phase transition boundary. When the chemical freeze-out line traverses the critical region of the first-order phase transition, the thermodynamic state of the system undergoes significant changes. Therefore, precise theoretical predictions of the first-order phase transition provide a foundational framework for experimental searches targeting the chemical freeze-out line and the CEP.

\begin{figure}
{\includegraphics[width=0.48\textwidth]{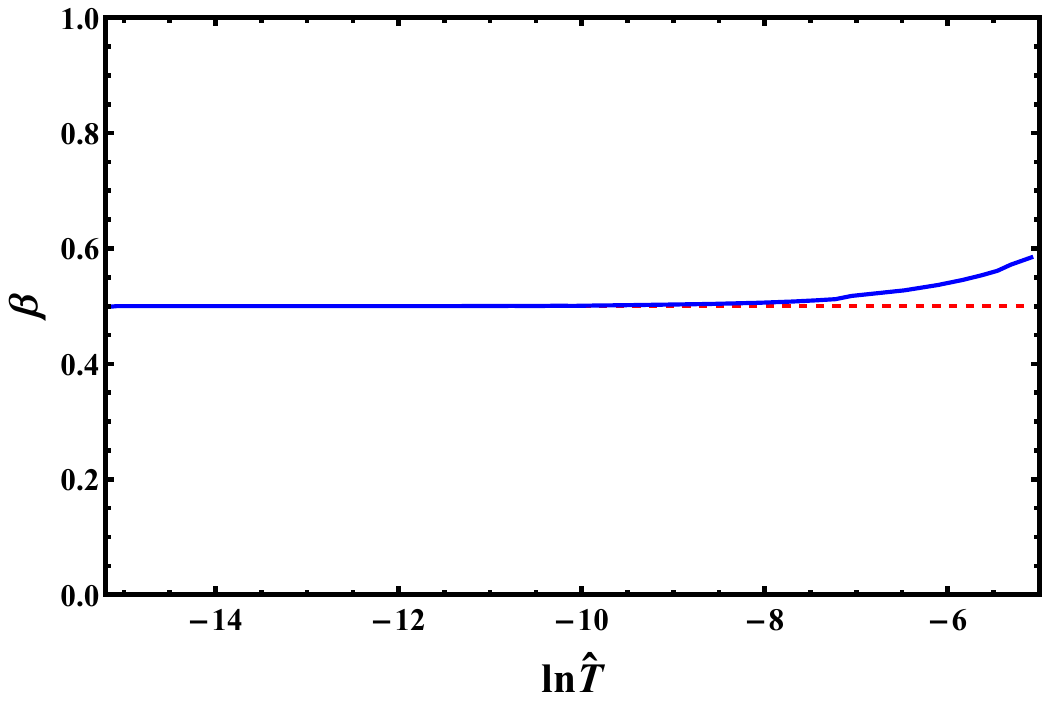}}
\caption[]{(Color online) The  critical exponent $\beta$ as a function of $\ln \hat{T}$ (blue line). The red dashed line represents the result of $\beta$ in mean field theory.}
\label{Tbeta.pdf}
\end{figure}

The critical exponent \(\beta\) is defined as follows:
\begin{eqnarray}
\Delta \rho \sim (T_c - T)^\beta,
\end{eqnarray}
where \(\Delta \rho = \rho_{>} - \rho_{<}\) (see Fig. \ref{phomjuTcaround} for the case where \(T < T_c\)). In this figure, the red curve intersects the blue straight line at three points. The left and right intersection points correspond to \(\rho_{<}\) and \(\rho_{>}\), respectively. In Fig. \ref{Tbeta.pdf}, we plot the critical exponent $\beta$ as a function of \(\ln \hat{T}\). The red dashed line represents the value of \(\beta\) predicted by mean field theory. Notably, $\beta$ converges as the temperature approaches \(T_{CEP}\). Moreover, near \(T_{CEP}\), the value of $\beta$ agrees with the mean field theory prediction for \(\beta\). However, as the temperature deviates from \(T_{CEP}\), the discrepancy between the slope and the mean field result increases significantly.

\begin{figure}
{\includegraphics[width=0.48\textwidth]{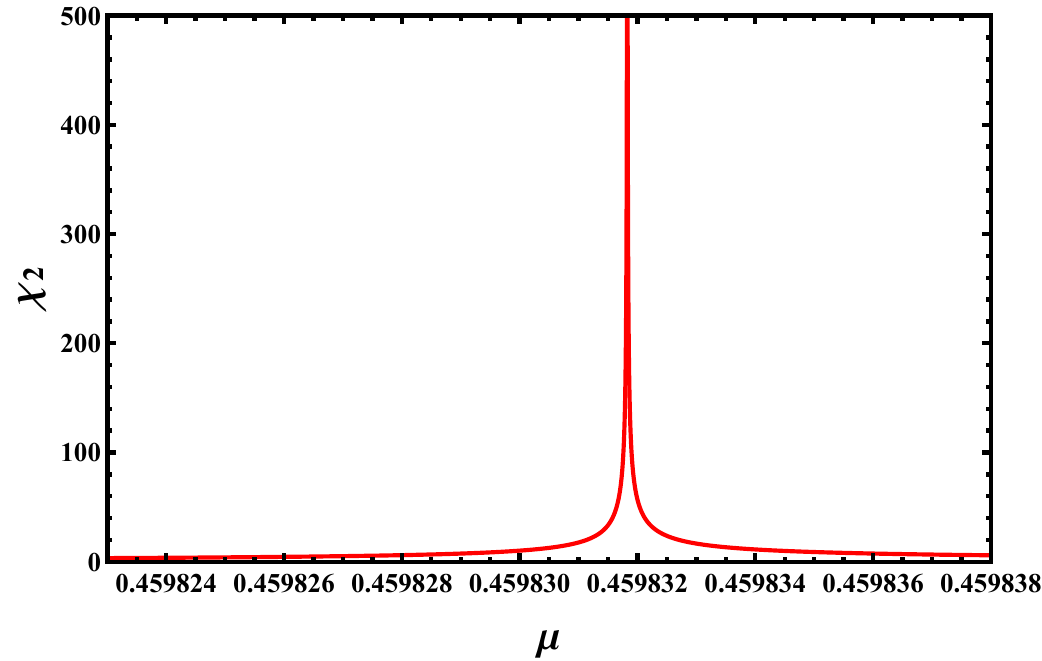}}
\caption[]{(Color online) The quark number susceptibility $\chi_2$ as a function of quark chemical potential $\mu$ around the CEP when $T=T_{CEP}$.}
\label{mjuchi2cep.pdf}
\end{figure}

\begin{figure}
{\includegraphics[width=0.48\textwidth]{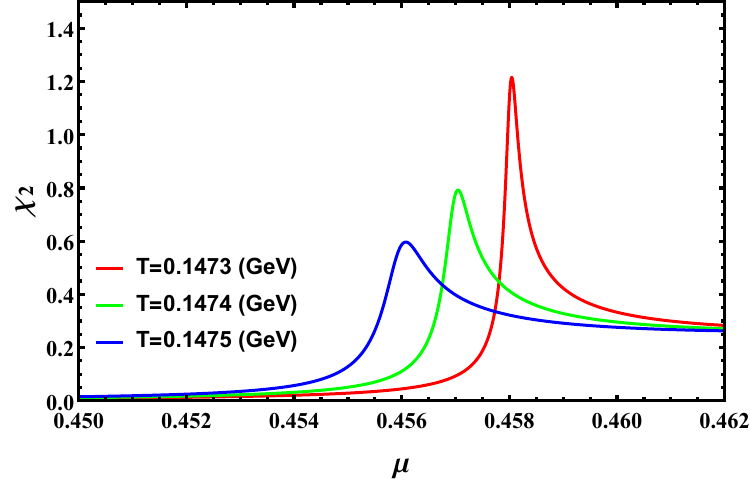}}
\caption[]{(Color online) The quark number susceptibility as a function of quark chemical potential for different temperatures when $T>T_{c}$.}
\label{mjuchi2.pdf}
\end{figure}

In Fig. \ref{mjuchi2cep.pdf}, we plot the quark number susceptibility $\chi_2$ as a function of the quark chemical potential $\mu$ at temperature \(T = T_{CEP}\). The figure reveals that the susceptibility exhibits a near-divergent behavior as the chemical potential approaches the critical value. In Fig. \ref{mjuchi2.pdf}, we present the quark number susceptibility as a function of the quark chemical potential for various temperatures above \(T_{CEP}\). Several characteristic behaviors are observed. As the quark chemical potential increases, the susceptibility generally follows a non-monotonic trend. At low chemical potentials, the susceptibility remains finite. However, as the chemical potential approaches the critical value, the susceptibility increases more rapidly, indicating an enhanced sensitivity of the quark number density to changes in the chemical potential. This behavior aligns with the expected critical phenomena near the critical point, where the system undergoes significant changes. Furthermore, for temperatures above \(T_c\), the rate of increase and the overall shape of the susceptibility curves vary. Higher temperatures result in a more gradual increase in susceptibility than lower temperatures. This temperature dependence suggests that thermal energy influences quark interactions and the system's response to changes in the chemical potential.

\begin{figure}
\subfigure[]
{\includegraphics[width=0.45\textwidth]{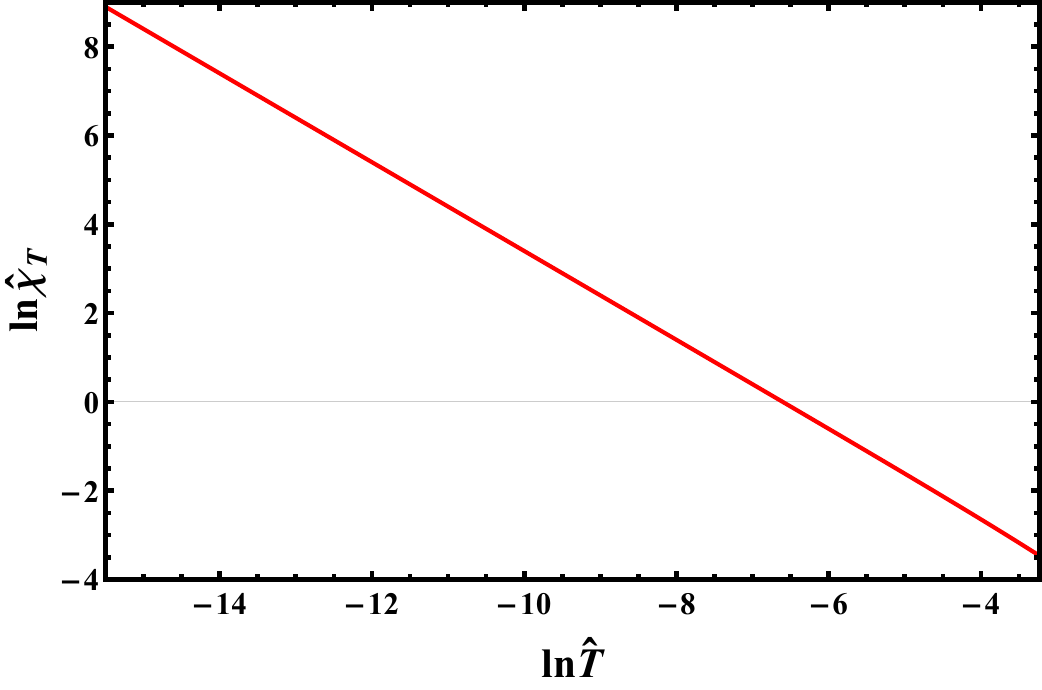}}
\subfigure[]
{\includegraphics[width=0.45\textwidth]{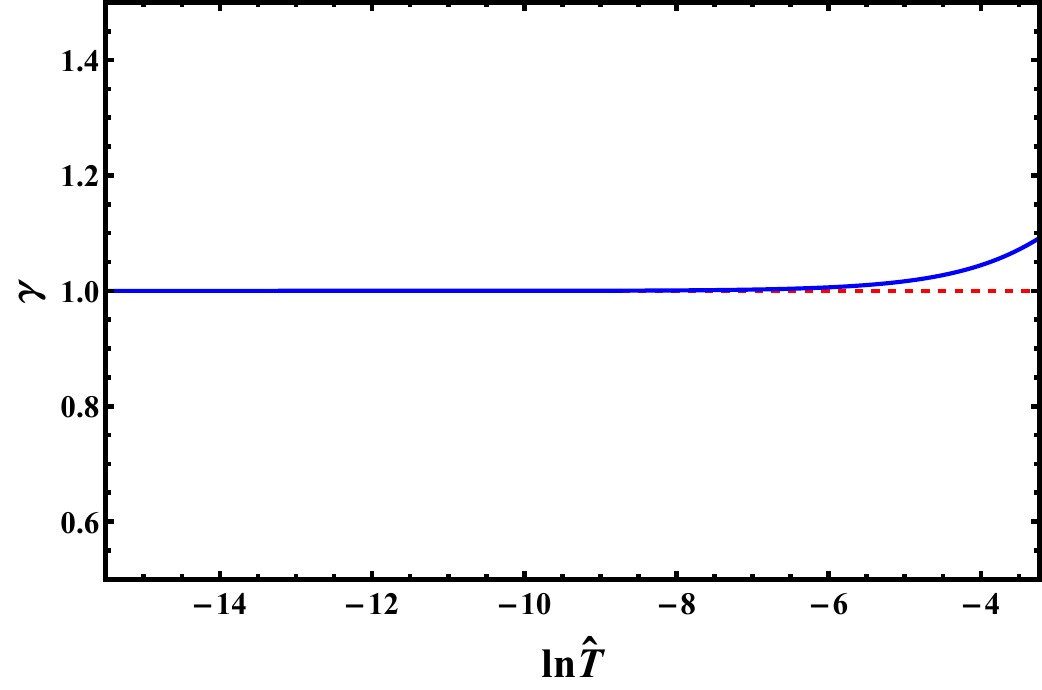}}
\caption[]{(Color online) (a) The reduced quark number susceptibility $\widehat{\chi}_{T}$ versus $\widehat{T}$ at fixed  baryon number density on a ln-ln plot. (b) The  critical exponent $\gamma$ as a function of the reduced temperature $\text{ln} \widehat{T}$. The dashed red line represents the value of $\gamma$ in the mean field.}
\label{Tgamma.pdf}
\end{figure}

The critical exponent \(\gamma\) is defined as:
\begin{eqnarray}
\chi_T \sim |T - T_{CEP}|^{-\gamma},
\end{eqnarray}
where \(\chi_T = \left( \frac{\partial \rho}{\partial \mu} \right)_T\). In Fig. \ref{Tgamma.pdf} (a), we show a ln - ln plot of the reduced quark number susceptibility \(\widehat{\chi}_T\) versus \(\widehat{T}\) at a fixed baryon number density. The corresponding $\beta$ is plotted as a function of the reduced temperature \(\ln \widehat{T}\) in Fig. \ref{Tgamma.pdf} (b). The dashed red line represents the value of \(\gamma\) predicted by mean field theory. A notable feature of the plot is the temperature-dependent slope, which decreases as the temperature increases. Given the sign difference between \(\gamma\) and the slope, it follows that \(\gamma\) increases with temperature. Importantly, when the system is far from the CEP, the value of \(\gamma\) can exceed 1. However, as the system approaches the CEP, \(\gamma\) converges to 1, consistent with the predictions of mean field theory.

Finally, the critical exponent \(\delta\) is defined at the critical isotherm \(T = T_{CEP}\) as follows:
\begin{eqnarray}
\frac{\rho(z_{h}, T_{CEP}) - \rho_{CEP}}{\rho_{CEP}} \sim (\frac{\mu(z_{h}, T_{CEP}) - \mu_{CEP}}{\mu_{CEP}})^{\frac{1}{\delta}},
\end{eqnarray}
where \(\mu - \mu_{CEP}\) and \(\rho - \rho_{CEP}\) represent the distances from the CEP. For simplicity, we define reduced baryon density \(\hat{\rho} = \frac{\rho(z_{h}, T_{CEP}) - \rho_{CEP}}{\rho_{CEP}}\) and quark chemical potential \(\hat{\mu} = \frac{\mu(z_{h}, T_{CEP}) - \mu_{CEP}}{\mu_{CEP}}\). By taking logarithms of both sides, we obtain \text{ln}\(\hat{\rho} = \frac{1}{\delta} \text{ln}\hat{\mu}\) or equivalently \text{ln}\(\hat{\mu} = \delta \text{ln}\hat{\rho}\). As pointed out in the introduction, it is interesting how the choice of direction approaching the CEP influences the computed values of critical exponents. Here, we take the critical exponent \(\delta\) as an example, when approaching the CEP from the opposite direction, it is important to note that both \(\mu - \mu_{CEP}\) and \(\rho - \rho_{CEP}\) are typically less than zero, therefore, their absolute values must be taken before applying logarithms. Correspondingly, \(\delta\) is defined by the relation \text{ln}\(|\hat{\rho}| \sim \text{ln}|\hat{\mu}|^{1/\delta}\) at \(T = T_{CEP}\).

\begin{figure}
\subfigure[]
{\includegraphics[width=0.45\textwidth]{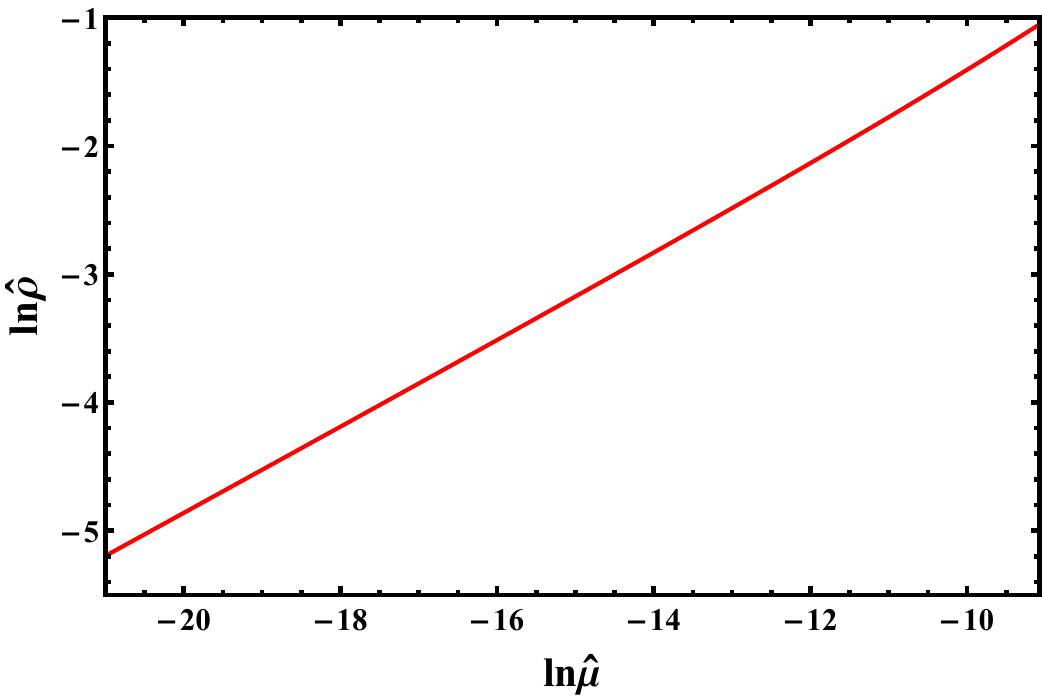}}
\subfigure[]
{\includegraphics[width=0.45\textwidth]{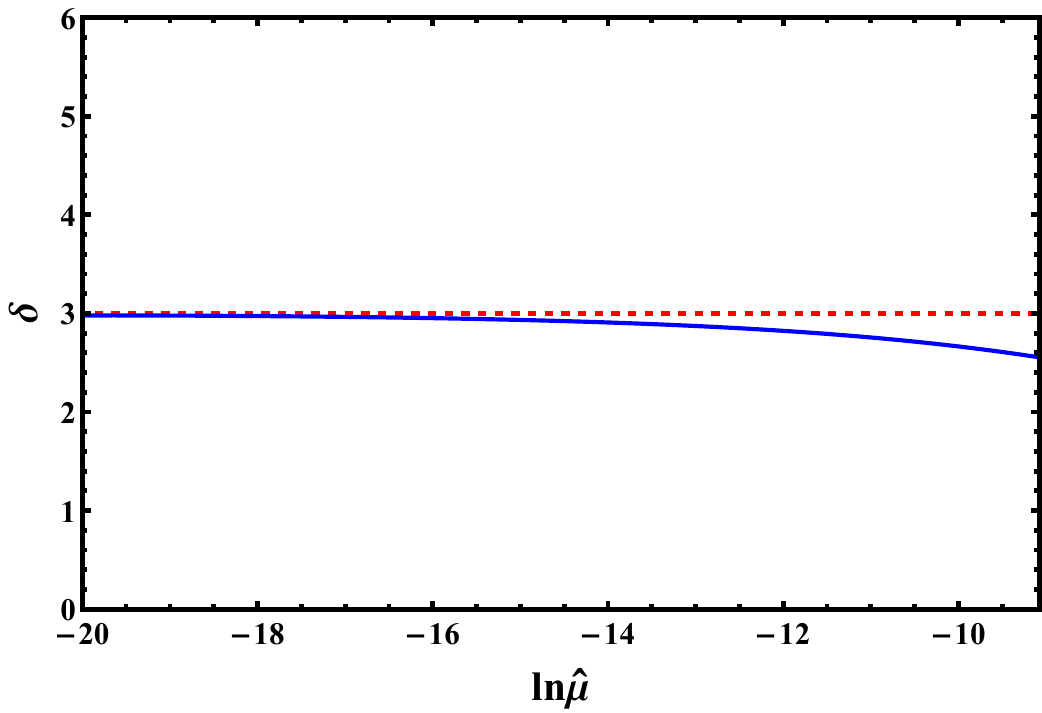}}
\caption[]{(Color online) (a) ln-ln plot of the reduced baryon density $\widehat{\rho}$ versus quark chemical potential $\widehat{\mu}$. (b) The  critical exponent $\delta$ versus  reduced quark chemical potential $\text{ln} \widehat{\mu}$. The red dashed line represents the result of $\delta$ in mean field.}
\label{mjupholeft.pdf}
\end{figure}

\begin{figure}
\subfigure[]
{\includegraphics[width=0.45\textwidth]{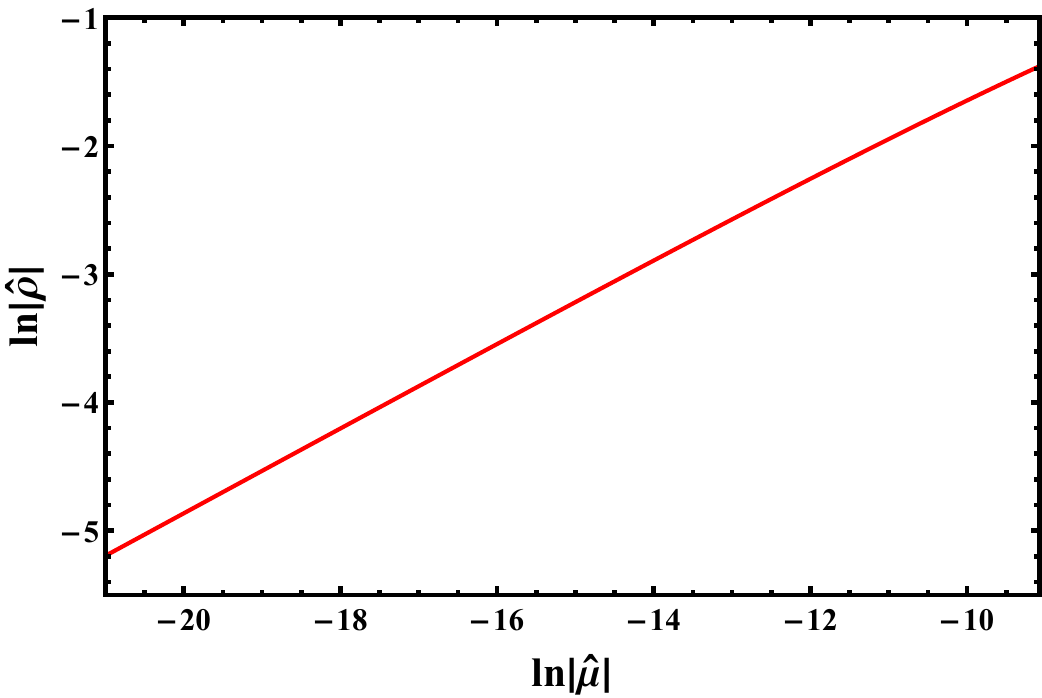}}
\subfigure[]
{\includegraphics[width=0.45\textwidth]{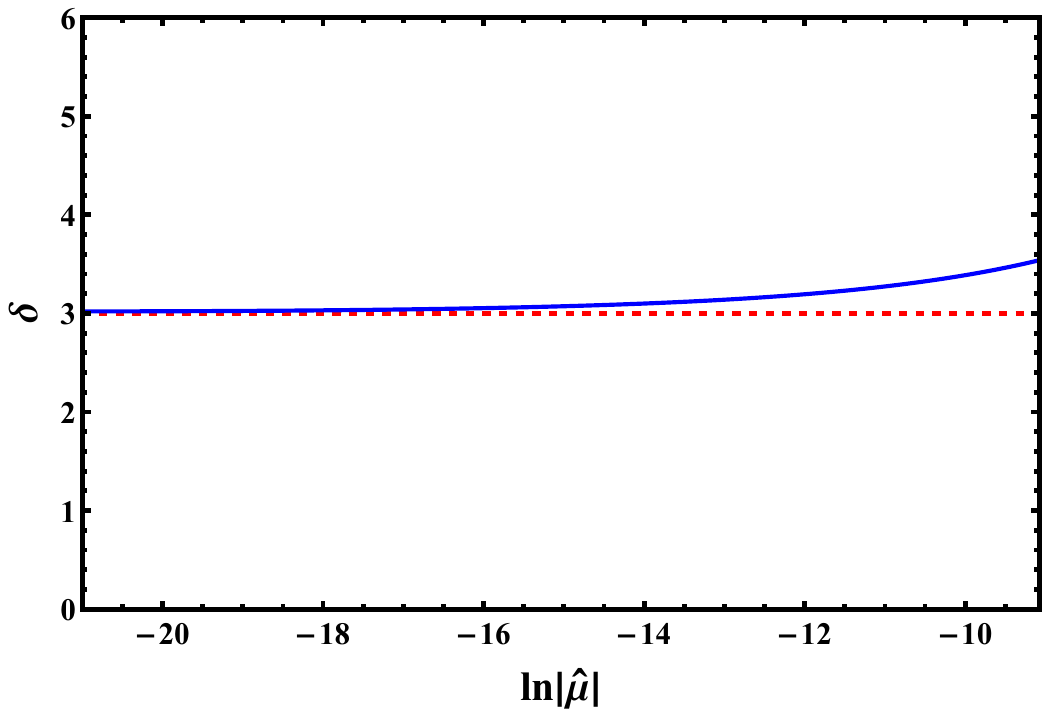}}
\caption[]{(Color online) (a)ln-ln plot of the reduced baryon density $\widehat{\rho}$ versus quark chemical potential $\widehat{\mu}$. Here, we approach the CEP from another direction. (b) The critical exponent $\delta$ versus the reduced quark chemical potential $\text{ln} \widehat{\mu}$. The red dashed line represents the result of $\delta$ in the mean field.}
\label{mjuphoright.pdf}
\end{figure}

In Figs. \ref{mjupholeft.pdf} and \ref{mjuphoright.pdf}, we present (a) a ln-ln plot of the reduced baryon density \(\widehat{\rho}\) versus the reduced quark chemical potential \(\widehat{\mu}\), and (b) the critical exponent \(\delta\) as a function of \(\ln \widehat{\mu}\). The red dashed line represents the critical exponent \(\delta\) predicted by mean field theory. When approaching the CEP from different directions, the initial values of \(\delta\) vary. As the system dynamically approaches the CEP, the values of \(\delta\) corresponding to different directions converge to a constant value of 3, consistent with mean field theory predictions. Therefore, if \(\delta\) is defined as the limiting value when \(z_h \rightarrow z_{h_{CEP}}\), then, the critical exponent \(\delta\) does not depend on whether the approach is from \(z_h > z_{h_{CEP}}\) or \(z_h < z_{h_{CEP}}\).
This study provides the first analysis of the critical exponent \(\delta\) near the CEP through different approach paths. Our results demonstrate that \(\delta\) converges to a universal value as the system approaches the CEP, suggesting a path-independent universal behavior. Furthermore, analogous universality is expected for other critical exponents when approaching the CEP from distinct approach paths.

These exponents are interrelated through scaling relations, which impose constraints between different critical exponents and reflect universality. Taking into account the errors in numerical calculations, our results are consistent with the following scaling relations: \(\alpha + 2\beta + \gamma = 2, \alpha + \beta(1 + \delta) = 2\).
It is important to note that even if the results satisfy the scaling relations, deviations from mean field theory warrant scrutiny. First, the precise location of the critical point is crucial, as critical exponents are highly sensitive to it. Second, the distance of the selected fitting region from the critical point significantly influences the critical exponents. Lastly, discrepancies may arise due to different paths taken when approaching the critical point. Therefore, in different holographic models, it is essential to conduct in-depth investigations to determine the precise location of the critical point, approach the CEP from various paths, and use limiting methods to ascertain the critical exponents.\\
\\

\begin{table}[ht!]
    \setlength{\tabcolsep}{0.5 mm}{
    \begin{tabular}{|c|c|c|c|c|}
     \cline{1-5}
      &    $\alpha$    &   $\beta$        & $\gamma$ & $\delta$ \\ \cline{1-5}
    Experiment & 0.110-0.116 &     0.316-0.327     &   1.23-1.25      &   4.6-4.9        \\ \cline{1-5}
    3D Ising  & 0.110(5) &  0.325$\pm$0.0015 & 1.2405$\pm$0.0015       & 4.82(4)    \\ \cline{1-5}
    Mean field  &  0  &  1/2 & 1          & 3          \\ \cline{1-5}
    DGR \cite{DeWolfe:2010he}  &  0   & 0.482          & 0.942          & 3.035   \\ \cline{1-5}
     hQCD \cite{Zhao:2023gur}   &   0.113   & 0.322          & 1.243          &  4.854   \\ \cline{1-5}
     hQCD \cite{Fu:2024wkn}  &    $1.97\times 10^{-7}$  & 0.5098          &  1.0159          &  3.0399   \\ \cline{1-5}
    hQCD(I) \cite{Cai:2024eqa}   &  0.002296  & 0.485518          &  0.9558187        & 3.00993  \\ \cline{1-5}
    hQCD(II) \cite{Cai:2024eqa}  &   0.001694  & 0.50373         &  0.91803        & 2.9455  \\ \cline{1-5}
    hQCD(III) \cite{Cai:2024eqa}   & 0.00917 & 0.3944          &  0.98696       & 3.9878   \\ \cline{1-5}
    Our results   & -0.000011 & 0.500113          &  0.999977       & 2.97991   \\ \cline{1-5}
    \end{tabular}}
\caption{Comparison of critical exponents from our model with experimental data from non-QCD fluids, 3D Ising numerical simulations, mean-field theory, and other holographic QCD approaches}
    \label{tab1}
\end{table}

Table \ref{tab1} compares the critical exponents derived from experiments on non-QCD fluids, the full quantum 3D Ising model, mean-field theory, various holographic QCD models, and the high-precision calculations from our model. Our numerical results show close agreement with the mean-field theoretical predictions. When approaching the CEP along a distinct thermodynamic path, the critical exponent \(\delta\) is calculated as \(3.02097\). For simplicity, this value is omitted from the table. The critical exponents are strongly sensitive to the location of the CEP \cite{Cai:2024eqa}, underscoring the necessity of high-precision determinations of the CEP. Our analyses demonstrate that deviations in the critical exponents grow with increasing \(\ln \widehat{T}\) or \(\ln \widehat{\mu}\). This trend highlights the critical role of precisely determining the CEP to ensure accurate calculations of these exponents.

\section{CONCLUSIONS\label{sec4}}
This paper presents a precise and rigorous calculation of the CEP, susceptibility, and critical exponents under finite temperature and chemical potential conditions. We systematically investigate the influence of the CEP's precise location, the choice of the fitting region, and the path taken to approach the CEP on determining critical exponents. Critical phenomena are characterized by power-law behaviors, which are inherently nonlinear. In previous studies, the determination of critical exponents has commonly relied on linear fitting, which inherently assumes that the slope variation is linear. However, the slope values obtained from linear fitting depend critically on both the location of the CEP and the fitting interval (i.e., the distance from the CEP). Currently, there is no standardized method for selecting this interval, leading to significant uncertainties in the determination of critical exponents. We propose that determining critical exponents by calculating the convergent value of the slope represents a more robust approach. However, this method also requires precise calculations of the CEP location to ensure accuracy. Our analysis reveals that the limiting method offers a more intuitive and comprehensive description of critical exponent behavior near the CEP than linear fitting. Linear fitting, while useful, is only applicable in regions extremely close to the CEP.

Furthermore, we demonstrate that different paths (or approaches from different directions) toward the CEP yield varying critical exponent values when the system is relatively far from the CEP. However, as the system approaches the CEP, these values converge to a universal value, indicating a universal behavior. Accurate predictions of the CEP's position and critical exponents not only deepen our understanding of the duality in AdS/CFT but also provide theoretical guidance for experimental searches for the CEP and for estimating the size of the critical region in the temperature direction \cite{Braun:2023qak}. These insights contribute to refining QCD theoretical models and advancing our understanding of QCD phase structure and critical phenomena.

In recent years, significant progress has been made in the study of the CEP and associated critical exponents in QCD, driven by major experimental initiatives. Facilities such as the RHIC at Brookhaven National Laboratory and the LHC at CERN have been instrumental in producing QGP through high-energy heavy ion collisions, yielding critical data on phase transition characteristics. Upcoming facilities, including the Facility for Antiproton and Ion Research (FAIR) in Germany and the Nuclotron-based Ion Collider Facility (NICA) in Russia, will further explore low- and medium-energy regions, enhancing our understanding of QCD phase transitions. Additionally, Japan's J-PARC is conducting experiments to study QCD behavior under high temperatures and densities. Collectively, these efforts are helping to unravel the complexities of strong interactions and offering valuable insights into the fundamental properties of matter under extreme conditions, which in turn deepens our understanding of critical phenomena in QCD.

In the context of AdS/CFT, achieving critical exponent results beyond mean-field approximations may require further refinement of the theoretical framework. For instance, incorporating critical phenomena of non-equilibrium phase transitions \cite{Matsumoto:2018ukk}, which align with Landau theory predictions, and introducing new physical mechanisms, such as scalar-vector interactions, could significantly influence critical exponents and phase diagrams. Another approach could be incorporating higher powers of the scalar potential and taking into account the temperature dependence of the dilaton field \cite {Chen:2018msc}. Additionally, QCD phase transitions under extreme conditions—such as finite temperature, baryon density, strong magnetic fields, and rapid rotation—are particularly interesting. Among these, rapid rotation has emerged as a key factor in studying phase transitions and the CEP \cite{Jiang:2016wvv,Wang:2018sur,Chen:2020ath,Chen:2021aiq,Fujimoto:2021xix,Zhao:2022uxc,Nunes:2024hzy,Chen:2024utf,Chen:2024jet,Chen:2024edy}. Correspondingly, the effects of fast rotation may also play a significant role in shaping critical exponents.

\section*{Acknowledgements}
We thank Danning Li and Yanqing Zhao for their useful discussions. The work has been supported by the National Natural Science Foundation of China (NSFC) under Grant Nos. 12375137, 12405154, and  12005114.

\bibliography{ref-lib}

\begin{thebibliography}{112}%
\makeatletter
\providecommand \@ifxundefined [1]{%
 \@ifx{#1\undefined}
}%
\providecommand \@ifnum [1]{%
 \ifnum #1\expandafter \@firstoftwo
 \else \expandafter \@secondoftwo
 \fi
}%
\providecommand \@ifx [1]{%
 \ifx #1\expandafter \@firstoftwo
 \else \expandafter \@secondoftwo
 \fi
}%
\providecommand \natexlab [1]{#1}%
\providecommand \enquote  [1]{``#1''}%
\providecommand \bibnamefont  [1]{#1}%
\providecommand \bibfnamefont [1]{#1}%
\providecommand \citenamefont [1]{#1}%
\providecommand \href@noop [0]{\@secondoftwo}%
\providecommand \href [0]{\begingroup \@sanitize@url \@href}%
\providecommand \@href[1]{\@@startlink{#1}\@@href}%
\providecommand \@@href[1]{\endgroup#1\@@endlink}%
\providecommand \@sanitize@url [0]{\catcode `\\12\catcode `\$12\catcode
  `\&12\catcode `\#12\catcode `\^12\catcode `\_12\catcode `\%12\relax}%
\providecommand \@@startlink[1]{}%
\providecommand \@@endlink[0]{}%
\providecommand \url  [0]{\begingroup\@sanitize@url \@url }%
\providecommand \@url [1]{\endgroup\@href {#1}{\urlprefix }}%
\providecommand \urlprefix  [0]{URL }%
\providecommand \Eprint [0]{\href }%
\providecommand \doibase [0]{http://dx.doi.org/}%
\providecommand \selectlanguage [0]{\@gobble}%
\providecommand \bibinfo  [0]{\@secondoftwo}%
\providecommand \bibfield  [0]{\@secondoftwo}%
\providecommand \translation [1]{[#1]}%
\providecommand \BibitemOpen [0]{}%
\providecommand \bibitemStop [0]{}%
\providecommand \bibitemNoStop [0]{.\EOS\space}%
\providecommand \EOS [0]{\spacefactor3000\relax}%
\providecommand \BibitemShut  [1]{\csname bibitem#1\endcsname}%
\let\auto@bib@innerbib\@empty
\bibitem [{\citenamefont {Espinosa}\ \emph {et~al.}(2010)\citenamefont
  {Espinosa}, \citenamefont {Konstandin}, \citenamefont {No},\ and\
  \citenamefont {Servant}}]{Espinosa:2010hh}%
  \BibitemOpen
  \bibfield  {author} {\bibinfo {author} {\bibfnamefont {J.~R.}\ \bibnamefont
  {Espinosa}}, \bibinfo {author} {\bibfnamefont {T.}~\bibnamefont
  {Konstandin}}, \bibinfo {author} {\bibfnamefont {J.~M.}\ \bibnamefont {No}},
  \ and\ \bibinfo {author} {\bibfnamefont {G.}~\bibnamefont {Servant}},\ }\href
  {\doibase 10.1088/1475-7516/2010/06/028} {\bibfield  {journal} {\bibinfo
  {journal} {JCAP}\ }\textbf {\bibinfo {volume} {06}},\ \bibinfo {pages} {028}
  (\bibinfo {year} {2010})},\ \Eprint {http://arxiv.org/abs/1004.4187}
  {arXiv:1004.4187 [hep-ph]} \BibitemShut {NoStop}%
\bibitem [{\citenamefont {Boeckel}\ and\ \citenamefont
  {Schaffner-Bielich}(2012)}]{Boeckel:2011yj}%
  \BibitemOpen
  \bibfield  {author} {\bibinfo {author} {\bibfnamefont {T.}~\bibnamefont
  {Boeckel}}\ and\ \bibinfo {author} {\bibfnamefont {J.}~\bibnamefont
  {Schaffner-Bielich}},\ }\href {\doibase 10.1103/PhysRevD.85.103506}
  {\bibfield  {journal} {\bibinfo  {journal} {Phys. Rev. D}\ }\textbf {\bibinfo
  {volume} {85}},\ \bibinfo {pages} {103506} (\bibinfo {year} {2012})},\
  \Eprint {http://arxiv.org/abs/1105.0832} {arXiv:1105.0832 [astro-ph.CO]}
  \BibitemShut {NoStop}%
\bibitem [{\citenamefont {Caprini}\ \emph {et~al.}(2016)\citenamefont {Caprini}
  \emph {et~al.}}]{Caprini:2015zlo}%
  \BibitemOpen
  \bibfield  {author} {\bibinfo {author} {\bibfnamefont {C.}~\bibnamefont
  {Caprini}} \emph {et~al.},\ }\href {\doibase 10.1088/1475-7516/2016/04/001}
  {\bibfield  {journal} {\bibinfo  {journal} {JCAP}\ }\textbf {\bibinfo
  {volume} {04}},\ \bibinfo {pages} {001} (\bibinfo {year} {2016})},\ \Eprint
  {http://arxiv.org/abs/1512.06239} {arXiv:1512.06239 [astro-ph.CO]}
  \BibitemShut {NoStop}%
\bibitem [{\citenamefont {Liu}\ \emph {et~al.}(2022)\citenamefont {Liu},
  \citenamefont {Bian}, \citenamefont {Cai}, \citenamefont {Guo},\ and\
  \citenamefont {Wang}}]{Liu:2021svg}%
  \BibitemOpen
  \bibfield  {author} {\bibinfo {author} {\bibfnamefont {J.}~\bibnamefont
  {Liu}}, \bibinfo {author} {\bibfnamefont {L.}~\bibnamefont {Bian}}, \bibinfo
  {author} {\bibfnamefont {R.-G.}\ \bibnamefont {Cai}}, \bibinfo {author}
  {\bibfnamefont {Z.-K.}\ \bibnamefont {Guo}}, \ and\ \bibinfo {author}
  {\bibfnamefont {S.-J.}\ \bibnamefont {Wang}},\ }\href {\doibase
  10.1103/PhysRevD.105.L021303} {\bibfield  {journal} {\bibinfo  {journal}
  {Phys. Rev. D}\ }\textbf {\bibinfo {volume} {105}},\ \bibinfo {pages}
  {L021303} (\bibinfo {year} {2022})},\ \Eprint
  {http://arxiv.org/abs/2106.05637} {arXiv:2106.05637 [astro-ph.CO]}
  \BibitemShut {NoStop}%
\bibitem [{\citenamefont {Shao}\ and\ \citenamefont
  {Huang}(2023)}]{Shao:2022oqw}%
  \BibitemOpen
  \bibfield  {author} {\bibinfo {author} {\bibfnamefont {J.}~\bibnamefont
  {Shao}}\ and\ \bibinfo {author} {\bibfnamefont {M.}~\bibnamefont {Huang}},\
  }\href {\doibase 10.1103/PhysRevD.107.043011} {\bibfield  {journal} {\bibinfo
   {journal} {Phys. Rev. D}\ }\textbf {\bibinfo {volume} {107}},\ \bibinfo
  {pages} {043011} (\bibinfo {year} {2023})},\ \Eprint
  {http://arxiv.org/abs/2209.13809} {arXiv:2209.13809 [hep-ph]} \BibitemShut
  {NoStop}%
\bibitem [{\citenamefont {Pasechnik}\ \emph {et~al.}(2024)\citenamefont
  {Pasechnik}, \citenamefont {Reichert}, \citenamefont {Sannino},\ and\
  \citenamefont {Wang}}]{Pasechnik:2023hwv}%
  \BibitemOpen
  \bibfield  {author} {\bibinfo {author} {\bibfnamefont {R.}~\bibnamefont
  {Pasechnik}}, \bibinfo {author} {\bibfnamefont {M.}~\bibnamefont {Reichert}},
  \bibinfo {author} {\bibfnamefont {F.}~\bibnamefont {Sannino}}, \ and\
  \bibinfo {author} {\bibfnamefont {Z.-W.}\ \bibnamefont {Wang}},\ }\href
  {\doibase 10.1007/JHEP02(2024)159} {\bibfield  {journal} {\bibinfo  {journal}
  {JHEP}\ }\textbf {\bibinfo {volume} {02}},\ \bibinfo {pages} {159} (\bibinfo
  {year} {2024})},\ \Eprint {http://arxiv.org/abs/2309.16755} {arXiv:2309.16755
  [hep-ph]} \BibitemShut {NoStop}%
\bibitem [{\citenamefont {Shao}\ \emph {et~al.}(2025)\citenamefont {Shao},
  \citenamefont {Mao},\ and\ \citenamefont {Huang}}]{Shao:2024dxt}%
  \BibitemOpen
  \bibfield  {author} {\bibinfo {author} {\bibfnamefont {J.}~\bibnamefont
  {Shao}}, \bibinfo {author} {\bibfnamefont {H.}~\bibnamefont {Mao}}, \ and\
  \bibinfo {author} {\bibfnamefont {M.}~\bibnamefont {Huang}},\ }\href
  {\doibase 10.1103/PhysRevD.111.023052} {\bibfield  {journal} {\bibinfo
  {journal} {Phys. Rev. D}\ }\textbf {\bibinfo {volume} {111}},\ \bibinfo
  {pages} {023052} (\bibinfo {year} {2025})},\ \Eprint
  {http://arxiv.org/abs/2410.06780} {arXiv:2410.06780 [hep-ph]} \BibitemShut
  {NoStop}%
\bibitem [{\citenamefont {Aoki}\ \emph {et~al.}(2006)\citenamefont {Aoki},
  \citenamefont {Endrodi}, \citenamefont {Fodor}, \citenamefont {Katz},\ and\
  \citenamefont {Szabo}}]{Aoki:2006we}%
  \BibitemOpen
  \bibfield  {author} {\bibinfo {author} {\bibfnamefont {Y.}~\bibnamefont
  {Aoki}}, \bibinfo {author} {\bibfnamefont {G.}~\bibnamefont {Endrodi}},
  \bibinfo {author} {\bibfnamefont {Z.}~\bibnamefont {Fodor}}, \bibinfo
  {author} {\bibfnamefont {S.~D.}\ \bibnamefont {Katz}}, \ and\ \bibinfo
  {author} {\bibfnamefont {K.~K.}\ \bibnamefont {Szabo}},\ }\href {\doibase
  10.1038/nature05120} {\bibfield  {journal} {\bibinfo  {journal} {Nature}\
  }\textbf {\bibinfo {volume} {443}},\ \bibinfo {pages} {675} (\bibinfo {year}
  {2006})},\ \Eprint {http://arxiv.org/abs/hep-lat/0611014}
  {arXiv:hep-lat/0611014} \BibitemShut {NoStop}%
\bibitem [{\citenamefont {Borsanyi}\ \emph {et~al.}(2014)\citenamefont
  {Borsanyi}, \citenamefont {Fodor}, \citenamefont {Hoelbling}, \citenamefont
  {Katz}, \citenamefont {Krieg},\ and\ \citenamefont
  {Szabo}}]{Borsanyi:2013bia}%
  \BibitemOpen
  \bibfield  {author} {\bibinfo {author} {\bibfnamefont {S.}~\bibnamefont
  {Borsanyi}}, \bibinfo {author} {\bibfnamefont {Z.}~\bibnamefont {Fodor}},
  \bibinfo {author} {\bibfnamefont {C.}~\bibnamefont {Hoelbling}}, \bibinfo
  {author} {\bibfnamefont {S.~D.}\ \bibnamefont {Katz}}, \bibinfo {author}
  {\bibfnamefont {S.}~\bibnamefont {Krieg}}, \ and\ \bibinfo {author}
  {\bibfnamefont {K.~K.}\ \bibnamefont {Szabo}},\ }\href {\doibase
  10.1016/j.physletb.2014.01.007} {\bibfield  {journal} {\bibinfo  {journal}
  {Phys. Lett. B}\ }\textbf {\bibinfo {volume} {730}},\ \bibinfo {pages} {99}
  (\bibinfo {year} {2014})},\ \Eprint {http://arxiv.org/abs/1309.5258}
  {arXiv:1309.5258 [hep-lat]} \BibitemShut {NoStop}%
\bibitem [{\citenamefont {Bazavov}\ \emph {et~al.}(2014)\citenamefont {Bazavov}
  \emph {et~al.}}]{HotQCD:2014kol}%
  \BibitemOpen
  \bibfield  {author} {\bibinfo {author} {\bibfnamefont {A.}~\bibnamefont
  {Bazavov}} \emph {et~al.} (\bibinfo {collaboration} {HotQCD}),\ }\href
  {\doibase 10.1103/PhysRevD.90.094503} {\bibfield  {journal} {\bibinfo
  {journal} {Phys. Rev. D}\ }\textbf {\bibinfo {volume} {90}},\ \bibinfo
  {pages} {094503} (\bibinfo {year} {2014})},\ \Eprint
  {http://arxiv.org/abs/1407.6387} {arXiv:1407.6387 [hep-lat]} \BibitemShut
  {NoStop}%
\bibitem [{\citenamefont {Bazavov}\ \emph {et~al.}(2019)\citenamefont {Bazavov}
  \emph {et~al.}}]{HotQCD:2018pds}%
  \BibitemOpen
  \bibfield  {author} {\bibinfo {author} {\bibfnamefont {A.}~\bibnamefont
  {Bazavov}} \emph {et~al.} (\bibinfo {collaboration} {HotQCD}),\ }\href
  {\doibase 10.1016/j.physletb.2019.05.013} {\bibfield  {journal} {\bibinfo
  {journal} {Phys. Lett. B}\ }\textbf {\bibinfo {volume} {795}},\ \bibinfo
  {pages} {15} (\bibinfo {year} {2019})},\ \Eprint
  {http://arxiv.org/abs/1812.08235} {arXiv:1812.08235 [hep-lat]} \BibitemShut
  {NoStop}%
\bibitem [{\citenamefont {Borsanyi}\ \emph {et~al.}(2020)\citenamefont
  {Borsanyi}, \citenamefont {Fodor}, \citenamefont {Guenther}, \citenamefont
  {Kara}, \citenamefont {Katz}, \citenamefont {Parotto}, \citenamefont
  {Pasztor}, \citenamefont {Ratti},\ and\ \citenamefont
  {Szabo}}]{Borsanyi:2020fev}%
  \BibitemOpen
  \bibfield  {author} {\bibinfo {author} {\bibfnamefont {S.}~\bibnamefont
  {Borsanyi}}, \bibinfo {author} {\bibfnamefont {Z.}~\bibnamefont {Fodor}},
  \bibinfo {author} {\bibfnamefont {J.~N.}\ \bibnamefont {Guenther}}, \bibinfo
  {author} {\bibfnamefont {R.}~\bibnamefont {Kara}}, \bibinfo {author}
  {\bibfnamefont {S.~D.}\ \bibnamefont {Katz}}, \bibinfo {author}
  {\bibfnamefont {P.}~\bibnamefont {Parotto}}, \bibinfo {author} {\bibfnamefont
  {A.}~\bibnamefont {Pasztor}}, \bibinfo {author} {\bibfnamefont
  {C.}~\bibnamefont {Ratti}}, \ and\ \bibinfo {author} {\bibfnamefont {K.~K.}\
  \bibnamefont {Szabo}},\ }\href {\doibase 10.1103/PhysRevLett.125.052001}
  {\bibfield  {journal} {\bibinfo  {journal} {Phys. Rev. Lett.}\ }\textbf
  {\bibinfo {volume} {125}},\ \bibinfo {pages} {052001} (\bibinfo {year}
  {2020})},\ \Eprint {http://arxiv.org/abs/2002.02821} {arXiv:2002.02821
  [hep-lat]} \BibitemShut {NoStop}%
\bibitem [{\citenamefont {Braguta}\ \emph {et~al.}(2021)\citenamefont
  {Braguta}, \citenamefont {Kotov}, \citenamefont {Kuznedelev},\ and\
  \citenamefont {Roenko}}]{Braguta:2021jgn}%
  \BibitemOpen
  \bibfield  {author} {\bibinfo {author} {\bibfnamefont {V.~V.}\ \bibnamefont
  {Braguta}}, \bibinfo {author} {\bibfnamefont {A.~Y.}\ \bibnamefont {Kotov}},
  \bibinfo {author} {\bibfnamefont {D.~D.}\ \bibnamefont {Kuznedelev}}, \ and\
  \bibinfo {author} {\bibfnamefont {A.~A.}\ \bibnamefont {Roenko}},\ }\href
  {\doibase 10.1103/PhysRevD.103.094515} {\bibfield  {journal} {\bibinfo
  {journal} {Phys. Rev. D}\ }\textbf {\bibinfo {volume} {103}},\ \bibinfo
  {pages} {094515} (\bibinfo {year} {2021})},\ \Eprint
  {http://arxiv.org/abs/2102.05084} {arXiv:2102.05084 [hep-lat]} \BibitemShut
  {NoStop}%
\bibitem [{\citenamefont {Braguta}\ \emph {et~al.}(2023)\citenamefont
  {Braguta}, \citenamefont {Kotov}, \citenamefont {Roenko},\ and\ \citenamefont
  {Sychev}}]{Braguta:2022str}%
  \BibitemOpen
  \bibfield  {author} {\bibinfo {author} {\bibfnamefont {V.~V.}\ \bibnamefont
  {Braguta}}, \bibinfo {author} {\bibfnamefont {A.}~\bibnamefont {Kotov}},
  \bibinfo {author} {\bibfnamefont {A.}~\bibnamefont {Roenko}}, \ and\ \bibinfo
  {author} {\bibfnamefont {D.}~\bibnamefont {Sychev}},\ }\href {\doibase
  10.22323/1.430.0190} {\bibfield  {journal} {\bibinfo  {journal} {PoS}\
  }\textbf {\bibinfo {volume} {LATTICE2022}},\ \bibinfo {pages} {190} (\bibinfo
  {year} {2023})},\ \Eprint {http://arxiv.org/abs/2212.03224} {arXiv:2212.03224
  [hep-lat]} \BibitemShut {NoStop}%
\bibitem [{\citenamefont {Yang}\ and\ \citenamefont
  {Huang}(2023)}]{Yang:2023vsw}%
  \BibitemOpen
  \bibfield  {author} {\bibinfo {author} {\bibfnamefont {J.-C.}\ \bibnamefont
  {Yang}}\ and\ \bibinfo {author} {\bibfnamefont {X.-G.}\ \bibnamefont
  {Huang}},\ }\href@noop {} {\  (\bibinfo {year} {2023})},\ \Eprint
  {http://arxiv.org/abs/2307.05755} {arXiv:2307.05755 [hep-lat]} \BibitemShut
  {NoStop}%
\bibitem [{\citenamefont {Braguta}\ \emph {et~al.}(2024)\citenamefont
  {Braguta}, \citenamefont {Chernodub}, \citenamefont {Gershtein},\ and\
  \citenamefont {Roenko}}]{Braguta:2024zpi}%
  \BibitemOpen
  \bibfield  {author} {\bibinfo {author} {\bibfnamefont {V.~V.}\ \bibnamefont
  {Braguta}}, \bibinfo {author} {\bibfnamefont {M.~N.}\ \bibnamefont
  {Chernodub}}, \bibinfo {author} {\bibfnamefont {Y.~A.}\ \bibnamefont
  {Gershtein}}, \ and\ \bibinfo {author} {\bibfnamefont {A.~A.}\ \bibnamefont
  {Roenko}},\ }\href@noop {} {\  (\bibinfo {year} {2024})},\ \Eprint
  {http://arxiv.org/abs/2411.15085} {arXiv:2411.15085 [hep-lat]} \BibitemShut
  {NoStop}%
\bibitem [{\citenamefont {Borsanyi}\ \emph {et~al.}(2025)\citenamefont
  {Borsanyi}, \citenamefont {Fodor}, \citenamefont {Guenther}, \citenamefont
  {Parotto}, \citenamefont {Pasztor}, \citenamefont {Ratti}, \citenamefont
  {Vovchenko},\ and\ \citenamefont {Wong}}]{Borsanyi:2025dyp}%
  \BibitemOpen
  \bibfield  {author} {\bibinfo {author} {\bibfnamefont {S.}~\bibnamefont
  {Borsanyi}}, \bibinfo {author} {\bibfnamefont {Z.}~\bibnamefont {Fodor}},
  \bibinfo {author} {\bibfnamefont {J.~N.}\ \bibnamefont {Guenther}}, \bibinfo
  {author} {\bibfnamefont {P.}~\bibnamefont {Parotto}}, \bibinfo {author}
  {\bibfnamefont {A.}~\bibnamefont {Pasztor}}, \bibinfo {author} {\bibfnamefont
  {C.}~\bibnamefont {Ratti}}, \bibinfo {author} {\bibfnamefont
  {V.}~\bibnamefont {Vovchenko}}, \ and\ \bibinfo {author} {\bibfnamefont
  {C.~H.}\ \bibnamefont {Wong}},\ }\href@noop {} {\  (\bibinfo {year}
  {2025})},\ \Eprint {http://arxiv.org/abs/2502.10267} {arXiv:2502.10267
  [hep-lat]} \BibitemShut {NoStop}%
\bibitem [{\citenamefont {Allton}\ \emph {et~al.}(2002)\citenamefont {Allton},
  \citenamefont {Ejiri}, \citenamefont {Hands}, \citenamefont {Kaczmarek},
  \citenamefont {Karsch}, \citenamefont {Laermann}, \citenamefont {Schmidt},\
  and\ \citenamefont {Scorzato}}]{Allton:2002zi}%
  \BibitemOpen
  \bibfield  {author} {\bibinfo {author} {\bibfnamefont {C.~R.}\ \bibnamefont
  {Allton}}, \bibinfo {author} {\bibfnamefont {S.}~\bibnamefont {Ejiri}},
  \bibinfo {author} {\bibfnamefont {S.~J.}\ \bibnamefont {Hands}}, \bibinfo
  {author} {\bibfnamefont {O.}~\bibnamefont {Kaczmarek}}, \bibinfo {author}
  {\bibfnamefont {F.}~\bibnamefont {Karsch}}, \bibinfo {author} {\bibfnamefont
  {E.}~\bibnamefont {Laermann}}, \bibinfo {author} {\bibfnamefont
  {C.}~\bibnamefont {Schmidt}}, \ and\ \bibinfo {author} {\bibfnamefont
  {L.}~\bibnamefont {Scorzato}},\ }\href {\doibase 10.1103/PhysRevD.66.074507}
  {\bibfield  {journal} {\bibinfo  {journal} {Phys. Rev. D}\ }\textbf {\bibinfo
  {volume} {66}},\ \bibinfo {pages} {074507} (\bibinfo {year} {2002})},\
  \Eprint {http://arxiv.org/abs/hep-lat/0204010} {arXiv:hep-lat/0204010}
  \BibitemShut {NoStop}%
\bibitem [{\citenamefont {Allton}\ \emph {et~al.}(2005)\citenamefont {Allton},
  \citenamefont {Doring}, \citenamefont {Ejiri}, \citenamefont {Hands},
  \citenamefont {Kaczmarek}, \citenamefont {Karsch}, \citenamefont {Laermann},\
  and\ \citenamefont {Redlich}}]{Allton:2005gk}%
  \BibitemOpen
  \bibfield  {author} {\bibinfo {author} {\bibfnamefont {C.~R.}\ \bibnamefont
  {Allton}}, \bibinfo {author} {\bibfnamefont {M.}~\bibnamefont {Doring}},
  \bibinfo {author} {\bibfnamefont {S.}~\bibnamefont {Ejiri}}, \bibinfo
  {author} {\bibfnamefont {S.~J.}\ \bibnamefont {Hands}}, \bibinfo {author}
  {\bibfnamefont {O.}~\bibnamefont {Kaczmarek}}, \bibinfo {author}
  {\bibfnamefont {F.}~\bibnamefont {Karsch}}, \bibinfo {author} {\bibfnamefont
  {E.}~\bibnamefont {Laermann}}, \ and\ \bibinfo {author} {\bibfnamefont
  {K.}~\bibnamefont {Redlich}},\ }\href {\doibase 10.1103/PhysRevD.71.054508}
  {\bibfield  {journal} {\bibinfo  {journal} {Phys. Rev. D}\ }\textbf {\bibinfo
  {volume} {71}},\ \bibinfo {pages} {054508} (\bibinfo {year} {2005})},\
  \Eprint {http://arxiv.org/abs/hep-lat/0501030} {arXiv:hep-lat/0501030}
  \BibitemShut {NoStop}%
\bibitem [{\citenamefont {Bazavov}\ \emph {et~al.}(2017)\citenamefont {Bazavov}
  \emph {et~al.}}]{Bazavov:2017dus}%
  \BibitemOpen
  \bibfield  {author} {\bibinfo {author} {\bibfnamefont {A.}~\bibnamefont
  {Bazavov}} \emph {et~al.},\ }\href {\doibase 10.1103/PhysRevD.95.054504}
  {\bibfield  {journal} {\bibinfo  {journal} {Phys. Rev. D}\ }\textbf {\bibinfo
  {volume} {95}},\ \bibinfo {pages} {054504} (\bibinfo {year} {2017})},\
  \Eprint {http://arxiv.org/abs/1701.04325} {arXiv:1701.04325 [hep-lat]}
  \BibitemShut {NoStop}%
\bibitem [{\citenamefont {Bors\'anyi}\ \emph {et~al.}(2021)\citenamefont
  {Bors\'anyi}, \citenamefont {Fodor}, \citenamefont {Guenther}, \citenamefont
  {Kara}, \citenamefont {Katz}, \citenamefont {Parotto}, \citenamefont
  {P\'asztor}, \citenamefont {Ratti},\ and\ \citenamefont
  {Szab\'o}}]{Borsanyi:2021sxv}%
  \BibitemOpen
  \bibfield  {author} {\bibinfo {author} {\bibfnamefont {S.}~\bibnamefont
  {Bors\'anyi}}, \bibinfo {author} {\bibfnamefont {Z.}~\bibnamefont {Fodor}},
  \bibinfo {author} {\bibfnamefont {J.~N.}\ \bibnamefont {Guenther}}, \bibinfo
  {author} {\bibfnamefont {R.}~\bibnamefont {Kara}}, \bibinfo {author}
  {\bibfnamefont {S.~D.}\ \bibnamefont {Katz}}, \bibinfo {author}
  {\bibfnamefont {P.}~\bibnamefont {Parotto}}, \bibinfo {author} {\bibfnamefont
  {A.}~\bibnamefont {P\'asztor}}, \bibinfo {author} {\bibfnamefont
  {C.}~\bibnamefont {Ratti}}, \ and\ \bibinfo {author} {\bibfnamefont {K.~K.}\
  \bibnamefont {Szab\'o}},\ }\href {\doibase 10.1103/PhysRevLett.126.232001}
  {\bibfield  {journal} {\bibinfo  {journal} {Phys. Rev. Lett.}\ }\textbf
  {\bibinfo {volume} {126}},\ \bibinfo {pages} {232001} (\bibinfo {year}
  {2021})},\ \Eprint {http://arxiv.org/abs/2102.06660} {arXiv:2102.06660
  [hep-lat]} \BibitemShut {NoStop}%
\bibitem [{\citenamefont {Hatsuda}\ and\ \citenamefont
  {Kunihiro}(1994)}]{Hatsuda1994}%
  \BibitemOpen
  \bibfield  {author} {\bibinfo {author} {\bibfnamefont {T.}~\bibnamefont
  {Hatsuda}}\ and\ \bibinfo {author} {\bibfnamefont {T.}~\bibnamefont
  {Kunihiro}},\ }\href@noop {} {\bibfield  {journal} {\bibinfo  {journal}
  {Physical Review Letters}\ }\textbf {\bibinfo {volume} {72}},\ \bibinfo
  {pages} {3505} (\bibinfo {year} {1994})}\BibitemShut {NoStop}%
\bibitem [{\citenamefont {Schwarz}\ \emph {et~al.}(1999)\citenamefont
  {Schwarz}, \citenamefont {Klevansky},\ and\ \citenamefont
  {Papp}}]{Schwarz:1999dj}%
  \BibitemOpen
  \bibfield  {author} {\bibinfo {author} {\bibfnamefont {T.~M.}\ \bibnamefont
  {Schwarz}}, \bibinfo {author} {\bibfnamefont {S.~P.}\ \bibnamefont
  {Klevansky}}, \ and\ \bibinfo {author} {\bibfnamefont {G.}~\bibnamefont
  {Papp}},\ }\href {\doibase 10.1103/PhysRevC.60.055205} {\bibfield  {journal}
  {\bibinfo  {journal} {Phys. Rev. C}\ }\textbf {\bibinfo {volume} {60}},\
  \bibinfo {pages} {055205} (\bibinfo {year} {1999})},\ \Eprint
  {http://arxiv.org/abs/nucl-th/9903048} {arXiv:nucl-th/9903048} \BibitemShut
  {NoStop}%
\bibitem [{\citenamefont {Zhuang}\ \emph {et~al.}(2000)\citenamefont {Zhuang},
  \citenamefont {Huang},\ and\ \citenamefont {Yang}}]{Zhuang:2000ub}%
  \BibitemOpen
  \bibfield  {author} {\bibinfo {author} {\bibfnamefont {P.}~\bibnamefont
  {Zhuang}}, \bibinfo {author} {\bibfnamefont {M.}~\bibnamefont {Huang}}, \
  and\ \bibinfo {author} {\bibfnamefont {Z.}~\bibnamefont {Yang}},\ }\href
  {\doibase 10.1103/PhysRevC.62.054901} {\bibfield  {journal} {\bibinfo
  {journal} {Phys. Rev. C}\ }\textbf {\bibinfo {volume} {62}},\ \bibinfo
  {pages} {054901} (\bibinfo {year} {2000})},\ \Eprint
  {http://arxiv.org/abs/nucl-th/0008043} {arXiv:nucl-th/0008043} \BibitemShut
  {NoStop}%
\bibitem [{\citenamefont {Barducci}\ \emph {et~al.}(2004)\citenamefont
  {Barducci}, \citenamefont {Casalbuoni}, \citenamefont {Pettini},\ and\
  \citenamefont {Ravagli}}]{Barducci:2004tt}%
  \BibitemOpen
  \bibfield  {author} {\bibinfo {author} {\bibfnamefont {A.}~\bibnamefont
  {Barducci}}, \bibinfo {author} {\bibfnamefont {R.}~\bibnamefont
  {Casalbuoni}}, \bibinfo {author} {\bibfnamefont {G.}~\bibnamefont {Pettini}},
  \ and\ \bibinfo {author} {\bibfnamefont {L.}~\bibnamefont {Ravagli}},\ }\href
  {\doibase 10.1103/PhysRevD.69.096004} {\bibfield  {journal} {\bibinfo
  {journal} {Phys. Rev. D}\ }\textbf {\bibinfo {volume} {69}},\ \bibinfo
  {pages} {096004} (\bibinfo {year} {2004})},\ \Eprint
  {http://arxiv.org/abs/hep-ph/0402104} {arXiv:hep-ph/0402104} \BibitemShut
  {NoStop}%
\bibitem [{\citenamefont {Fukushima}(2004)}]{Fukushima2004}%
  \BibitemOpen
  \bibfield  {author} {\bibinfo {author} {\bibfnamefont {K.}~\bibnamefont
  {Fukushima}},\ }\href@noop {} {\bibfield  {journal} {\bibinfo  {journal}
  {Physical Review D}\ }\textbf {\bibinfo {volume} {70}},\ \bibinfo {pages}
  {014013} (\bibinfo {year} {2004})}\BibitemShut {NoStop}%
\bibitem [{\citenamefont {C.~Ratti}\ and\ \citenamefont
  {Tandy}(2009)}]{Ratti2009}%
  \BibitemOpen
  \bibfield  {author} {\bibinfo {author} {\bibfnamefont {R.~T.}\ \bibnamefont
  {C.~Ratti}}\ and\ \bibinfo {author} {\bibfnamefont {P.}~\bibnamefont
  {Tandy}},\ }\href@noop {} {\bibfield  {journal} {\bibinfo  {journal}
  {Physical Review C}\ }\textbf {\bibinfo {volume} {80}},\ \bibinfo {pages}
  {055202} (\bibinfo {year} {2009})}\BibitemShut {NoStop}%
\bibitem [{\citenamefont {Brauner}\ and\ \citenamefont
  {Matsuura}(2010)}]{Brauner2010}%
  \BibitemOpen
  \bibfield  {author} {\bibinfo {author} {\bibfnamefont {T.}~\bibnamefont
  {Brauner}}\ and\ \bibinfo {author} {\bibfnamefont {T.}~\bibnamefont
  {Matsuura}},\ }\href@noop {} {\bibfield  {journal} {\bibinfo  {journal}
  {Journal of Physics G: Nuclear and Particle Physics}\ }\textbf {\bibinfo
  {volume} {37}},\ \bibinfo {pages} {045032} (\bibinfo {year}
  {2010})}\BibitemShut {NoStop}%
\bibitem [{\citenamefont {Sasaki}\ and\ \citenamefont
  {Nara}(2011)}]{Sasaki2011}%
  \BibitemOpen
  \bibfield  {author} {\bibinfo {author} {\bibfnamefont {K.}~\bibnamefont
  {Sasaki}}\ and\ \bibinfo {author} {\bibfnamefont {Y.}~\bibnamefont {Nara}},\
  }\href@noop {} {\bibfield  {journal} {\bibinfo  {journal} {Physical Review
  C}\ }\textbf {\bibinfo {volume} {83}},\ \bibinfo {pages} {015205} (\bibinfo
  {year} {2011})}\BibitemShut {NoStop}%
\bibitem [{\citenamefont {Mao}\ and\ \citenamefont {Zhang}(2013)}]{Mao2013}%
  \BibitemOpen
  \bibfield  {author} {\bibinfo {author} {\bibfnamefont {S.}~\bibnamefont
  {Mao}}\ and\ \bibinfo {author} {\bibfnamefont {Y.}~\bibnamefont {Zhang}},\
  }\href@noop {} {\bibfield  {journal} {\bibinfo  {journal} {Physica A:
  Statistical Mechanics and its Applications}\ }\textbf {\bibinfo {volume}
  {392}},\ \bibinfo {pages} {782} (\bibinfo {year} {2013})}\BibitemShut
  {NoStop}%
\bibitem [{\citenamefont {Jiang}\ and\ \citenamefont
  {Liao}(2016)}]{Jiang:2016wvv}%
  \BibitemOpen
  \bibfield  {author} {\bibinfo {author} {\bibfnamefont {Y.}~\bibnamefont
  {Jiang}}\ and\ \bibinfo {author} {\bibfnamefont {J.}~\bibnamefont {Liao}},\
  }\href {\doibase 10.1103/PhysRevLett.117.192302} {\bibfield  {journal}
  {\bibinfo  {journal} {Phys. Rev. Lett.}\ }\textbf {\bibinfo {volume} {117}},\
  \bibinfo {pages} {192302} (\bibinfo {year} {2016})},\ \Eprint
  {http://arxiv.org/abs/1606.03808} {arXiv:1606.03808 [hep-ph]} \BibitemShut
  {NoStop}%
\bibitem [{\citenamefont {Sun}\ \emph {et~al.}(2021)\citenamefont {Sun},
  \citenamefont {Ko}, \citenamefont {Cao},\ and\ \citenamefont
  {Li}}]{Sun:2020bbn}%
  \BibitemOpen
  \bibfield  {author} {\bibinfo {author} {\bibfnamefont {K.-J.}\ \bibnamefont
  {Sun}}, \bibinfo {author} {\bibfnamefont {C.-M.}\ \bibnamefont {Ko}},
  \bibinfo {author} {\bibfnamefont {S.}~\bibnamefont {Cao}}, \ and\ \bibinfo
  {author} {\bibfnamefont {F.}~\bibnamefont {Li}},\ }\href {\doibase
  10.1103/PhysRevD.103.014006} {\bibfield  {journal} {\bibinfo  {journal}
  {Phys. Rev. D}\ }\textbf {\bibinfo {volume} {103}},\ \bibinfo {pages}
  {014006} (\bibinfo {year} {2021})},\ \Eprint
  {http://arxiv.org/abs/2004.05754} {arXiv:2004.05754 [nucl-th]} \BibitemShut
  {NoStop}%
\bibitem [{\citenamefont {Sun}\ \emph {et~al.}(2023{\natexlab{a}})\citenamefont
  {Sun}, \citenamefont {Li}, \citenamefont {Wen}, \citenamefont {Huang},\ and\
  \citenamefont {Xie}}]{Sun:2023yux}%
  \BibitemOpen
  \bibfield  {author} {\bibinfo {author} {\bibfnamefont {F.}~\bibnamefont
  {Sun}}, \bibinfo {author} {\bibfnamefont {S.}~\bibnamefont {Li}}, \bibinfo
  {author} {\bibfnamefont {R.}~\bibnamefont {Wen}}, \bibinfo {author}
  {\bibfnamefont {A.}~\bibnamefont {Huang}}, \ and\ \bibinfo {author}
  {\bibfnamefont {W.}~\bibnamefont {Xie}},\ }\href@noop {} {\  (\bibinfo {year}
  {2023}{\natexlab{a}})},\ \Eprint {http://arxiv.org/abs/2310.18942}
  {arXiv:2310.18942 [hep-ph]} \BibitemShut {NoStop}%
\bibitem [{\citenamefont {Qiu}\ \emph {et~al.}(2023)\citenamefont {Qiu},
  \citenamefont {Feng},\ and\ \citenamefont {Zhu}}]{Qiu:2023ezo}%
  \BibitemOpen
  \bibfield  {author} {\bibinfo {author} {\bibfnamefont {Y.-W.}\ \bibnamefont
  {Qiu}}, \bibinfo {author} {\bibfnamefont {S.-Q.}\ \bibnamefont {Feng}}, \
  and\ \bibinfo {author} {\bibfnamefont {X.-Q.}\ \bibnamefont {Zhu}},\ }\href
  {\doibase 10.1103/PhysRevD.108.116022} {\bibfield  {journal} {\bibinfo
  {journal} {Phys. Rev. D}\ }\textbf {\bibinfo {volume} {108}},\ \bibinfo
  {pages} {116022} (\bibinfo {year} {2023})},\ \Eprint
  {http://arxiv.org/abs/2307.13193} {arXiv:2307.13193 [hep-ph]} \BibitemShut
  {NoStop}%
\bibitem [{\citenamefont {Bao}\ and\ \citenamefont {Feng}(2024)}]{Bao:2024glw}%
  \BibitemOpen
  \bibfield  {author} {\bibinfo {author} {\bibfnamefont {Y.-R.}\ \bibnamefont
  {Bao}}\ and\ \bibinfo {author} {\bibfnamefont {S.-Q.}\ \bibnamefont {Feng}},\
  }\href {\doibase 10.1103/PhysRevD.109.096033} {\bibfield  {journal} {\bibinfo
   {journal} {Phys. Rev. D}\ }\textbf {\bibinfo {volume} {109}},\ \bibinfo
  {pages} {096033} (\bibinfo {year} {2024})},\ \Eprint
  {http://arxiv.org/abs/2403.16541} {arXiv:2403.16541 [hep-ph]} \BibitemShut
  {NoStop}%
\bibitem [{\citenamefont {Hua}\ and\ \citenamefont {Feng}(2024)}]{Hua:2024bwn}%
  \BibitemOpen
  \bibfield  {author} {\bibinfo {author} {\bibfnamefont {Y.}~\bibnamefont
  {Hua}}\ and\ \bibinfo {author} {\bibfnamefont {S.-Q.}\ \bibnamefont {Feng}},\
  }\href@noop {} {\  (\bibinfo {year} {2024})},\ \Eprint
  {http://arxiv.org/abs/2412.06398} {arXiv:2412.06398 [hep-ph]} \BibitemShut
  {NoStop}%
\bibitem [{\citenamefont {Klevansky}(1992)}]{Klevansky:1992qe}%
  \BibitemOpen
  \bibfield  {author} {\bibinfo {author} {\bibfnamefont {S.~P.}\ \bibnamefont
  {Klevansky}},\ }\href {\doibase 10.1103/RevModPhys.64.649} {\bibfield
  {journal} {\bibinfo  {journal} {Rev. Mod. Phys.}\ }\textbf {\bibinfo {volume}
  {64}},\ \bibinfo {pages} {649} (\bibinfo {year} {1992})}\BibitemShut
  {NoStop}%
\bibitem [{\citenamefont {Buballa}(2005)}]{Buballa:2003qv}%
  \BibitemOpen
  \bibfield  {author} {\bibinfo {author} {\bibfnamefont {M.}~\bibnamefont
  {Buballa}},\ }\href {\doibase 10.1016/j.physrep.2004.11.004} {\bibfield
  {journal} {\bibinfo  {journal} {Phys. Rept.}\ }\textbf {\bibinfo {volume}
  {407}},\ \bibinfo {pages} {205} (\bibinfo {year} {2005})},\ \Eprint
  {http://arxiv.org/abs/hep-ph/0402234} {arXiv:hep-ph/0402234} \BibitemShut
  {NoStop}%
\bibitem [{\citenamefont {Kohyama}\ \emph {et~al.}(2015)\citenamefont
  {Kohyama}, \citenamefont {Kimura},\ and\ \citenamefont
  {Inagaki}}]{Kohyama:2015hix}%
  \BibitemOpen
  \bibfield  {author} {\bibinfo {author} {\bibfnamefont {H.}~\bibnamefont
  {Kohyama}}, \bibinfo {author} {\bibfnamefont {D.}~\bibnamefont {Kimura}}, \
  and\ \bibinfo {author} {\bibfnamefont {T.}~\bibnamefont {Inagaki}},\ }\href
  {\doibase 10.1016/j.nuclphysb.2015.05.015} {\bibfield  {journal} {\bibinfo
  {journal} {Nucl. Phys. B}\ }\textbf {\bibinfo {volume} {896}},\ \bibinfo
  {pages} {682} (\bibinfo {year} {2015})},\ \Eprint
  {http://arxiv.org/abs/1501.00449} {arXiv:1501.00449 [hep-ph]} \BibitemShut
  {NoStop}%
\bibitem [{\citenamefont {Roberts}\ and\ \citenamefont
  {Williams}(1994)}]{Roberts:1994dr}%
  \BibitemOpen
  \bibfield  {author} {\bibinfo {author} {\bibfnamefont {C.~D.}\ \bibnamefont
  {Roberts}}\ and\ \bibinfo {author} {\bibfnamefont {A.~G.}\ \bibnamefont
  {Williams}},\ }\href {\doibase 10.1016/0146-6410(94)90049-3} {\bibfield
  {journal} {\bibinfo  {journal} {Prog. Part. Nucl. Phys.}\ }\textbf {\bibinfo
  {volume} {33}},\ \bibinfo {pages} {477} (\bibinfo {year} {1994})},\ \Eprint
  {http://arxiv.org/abs/hep-ph/9403224} {arXiv:hep-ph/9403224} \BibitemShut
  {NoStop}%
\bibitem [{\citenamefont {Alkofer}\ and\ \citenamefont {von
  Smekal}(2001)}]{Alkofer:2000wg}%
  \BibitemOpen
  \bibfield  {author} {\bibinfo {author} {\bibfnamefont {R.}~\bibnamefont
  {Alkofer}}\ and\ \bibinfo {author} {\bibfnamefont {L.}~\bibnamefont {von
  Smekal}},\ }\href {\doibase 10.1016/S0370-1573(01)00010-2} {\bibfield
  {journal} {\bibinfo  {journal} {Phys. Rept.}\ }\textbf {\bibinfo {volume}
  {353}},\ \bibinfo {pages} {281} (\bibinfo {year} {2001})},\ \Eprint
  {http://arxiv.org/abs/hep-ph/0007355} {arXiv:hep-ph/0007355} \BibitemShut
  {NoStop}%
\bibitem [{\citenamefont {Cloet}\ and\ \citenamefont
  {Roberts}(2014)}]{Cloet:2013jya}%
  \BibitemOpen
  \bibfield  {author} {\bibinfo {author} {\bibfnamefont {I.~C.}\ \bibnamefont
  {Cloet}}\ and\ \bibinfo {author} {\bibfnamefont {C.~D.}\ \bibnamefont
  {Roberts}},\ }\href {\doibase 10.1016/j.ppnp.2014.02.001} {\bibfield
  {journal} {\bibinfo  {journal} {Prog. Part. Nucl. Phys.}\ }\textbf {\bibinfo
  {volume} {77}},\ \bibinfo {pages} {1} (\bibinfo {year} {2014})},\ \Eprint
  {http://arxiv.org/abs/1310.2651} {arXiv:1310.2651 [nucl-th]} \BibitemShut
  {NoStop}%
\bibitem [{\citenamefont {Chen}\ \emph {et~al.}(2015)\citenamefont {Chen},
  \citenamefont {Deng},\ and\ \citenamefont {Labun}}]{Chen:2014ufa}%
  \BibitemOpen
  \bibfield  {author} {\bibinfo {author} {\bibfnamefont {J.-W.}\ \bibnamefont
  {Chen}}, \bibinfo {author} {\bibfnamefont {J.}~\bibnamefont {Deng}}, \ and\
  \bibinfo {author} {\bibfnamefont {L.}~\bibnamefont {Labun}},\ }\href
  {\doibase 10.1103/PhysRevD.92.054019} {\bibfield  {journal} {\bibinfo
  {journal} {Phys. Rev. D}\ }\textbf {\bibinfo {volume} {92}},\ \bibinfo
  {pages} {054019} (\bibinfo {year} {2015})},\ \Eprint
  {http://arxiv.org/abs/1410.5454} {arXiv:1410.5454 [hep-ph]} \BibitemShut
  {NoStop}%
\bibitem [{\citenamefont {Chen}\ \emph {et~al.}(2016)\citenamefont {Chen},
  \citenamefont {Deng}, \citenamefont {Kohyama},\ and\ \citenamefont
  {Labun}}]{Chen:2015dra}%
  \BibitemOpen
  \bibfield  {author} {\bibinfo {author} {\bibfnamefont {J.-W.}\ \bibnamefont
  {Chen}}, \bibinfo {author} {\bibfnamefont {J.}~\bibnamefont {Deng}}, \bibinfo
  {author} {\bibfnamefont {H.}~\bibnamefont {Kohyama}}, \ and\ \bibinfo
  {author} {\bibfnamefont {L.}~\bibnamefont {Labun}},\ }\href {\doibase
  10.1103/PhysRevD.93.034037} {\bibfield  {journal} {\bibinfo  {journal} {Phys.
  Rev. D}\ }\textbf {\bibinfo {volume} {93}},\ \bibinfo {pages} {034037}
  (\bibinfo {year} {2016})},\ \Eprint {http://arxiv.org/abs/1509.04968}
  {arXiv:1509.04968 [hep-ph]} \BibitemShut {NoStop}%
\bibitem [{\citenamefont {Fan}\ \emph {et~al.}(2017)\citenamefont {Fan},
  \citenamefont {Luo},\ and\ \citenamefont {Zong}}]{Fan:2016ovc}%
  \BibitemOpen
  \bibfield  {author} {\bibinfo {author} {\bibfnamefont {W.}~\bibnamefont
  {Fan}}, \bibinfo {author} {\bibfnamefont {X.}~\bibnamefont {Luo}}, \ and\
  \bibinfo {author} {\bibfnamefont {H.-S.}\ \bibnamefont {Zong}},\ }\href
  {\doibase 10.1142/S0217751X17500610} {\bibfield  {journal} {\bibinfo
  {journal} {Int. J. Mod. Phys. A}\ }\textbf {\bibinfo {volume} {32}},\
  \bibinfo {pages} {1750061} (\bibinfo {year} {2017})},\ \Eprint
  {http://arxiv.org/abs/1608.07903} {arXiv:1608.07903 [hep-ph]} \BibitemShut
  {NoStop}%
\bibitem [{\citenamefont {Fan}\ \emph {et~al.}(2019)\citenamefont {Fan},
  \citenamefont {Luo},\ and\ \citenamefont {Zong}}]{Fan:2017mrk}%
  \BibitemOpen
  \bibfield  {author} {\bibinfo {author} {\bibfnamefont {W.}~\bibnamefont
  {Fan}}, \bibinfo {author} {\bibfnamefont {X.}~\bibnamefont {Luo}}, \ and\
  \bibinfo {author} {\bibfnamefont {H.}~\bibnamefont {Zong}},\ }\href {\doibase
  10.1088/1674-1137/43/3/033103} {\bibfield  {journal} {\bibinfo  {journal}
  {Chin. Phys. C}\ }\textbf {\bibinfo {volume} {43}},\ \bibinfo {pages}
  {033103} (\bibinfo {year} {2019})},\ \Eprint
  {http://arxiv.org/abs/1702.08674} {arXiv:1702.08674 [hep-ph]} \BibitemShut
  {NoStop}%
\bibitem [{\citenamefont {Fu}\ and\ \citenamefont {Wu}(2010)}]{Fu:2010ay}%
  \BibitemOpen
  \bibfield  {author} {\bibinfo {author} {\bibfnamefont {W.-j.}\ \bibnamefont
  {Fu}}\ and\ \bibinfo {author} {\bibfnamefont {Y.-l.}\ \bibnamefont {Wu}},\
  }\href {\doibase 10.1103/PhysRevD.82.074013} {\bibfield  {journal} {\bibinfo
  {journal} {Phys. Rev. D}\ }\textbf {\bibinfo {volume} {82}},\ \bibinfo
  {pages} {074013} (\bibinfo {year} {2010})},\ \Eprint
  {http://arxiv.org/abs/1008.3684} {arXiv:1008.3684 [hep-ph]} \BibitemShut
  {NoStop}%
\bibitem [{\citenamefont {Bowman}\ and\ \citenamefont
  {Kapusta}(2009)}]{Bowman:2008kc}%
  \BibitemOpen
  \bibfield  {author} {\bibinfo {author} {\bibfnamefont {E.~S.}\ \bibnamefont
  {Bowman}}\ and\ \bibinfo {author} {\bibfnamefont {J.~I.}\ \bibnamefont
  {Kapusta}},\ }\href {\doibase 10.1103/PhysRevC.79.015202} {\bibfield
  {journal} {\bibinfo  {journal} {Phys. Rev. C}\ }\textbf {\bibinfo {volume}
  {79}},\ \bibinfo {pages} {015202} (\bibinfo {year} {2009})},\ \Eprint
  {http://arxiv.org/abs/0810.0042} {arXiv:0810.0042 [nucl-th]} \BibitemShut
  {NoStop}%
\bibitem [{\citenamefont {Schaefer}\ and\ \citenamefont
  {Wambach}(2005)}]{Schaefer:2004en}%
  \BibitemOpen
  \bibfield  {author} {\bibinfo {author} {\bibfnamefont {B.-J.}\ \bibnamefont
  {Schaefer}}\ and\ \bibinfo {author} {\bibfnamefont {J.}~\bibnamefont
  {Wambach}},\ }\href {\doibase 10.1016/j.nuclphysa.2005.04.012} {\bibfield
  {journal} {\bibinfo  {journal} {Nucl. Phys. A}\ }\textbf {\bibinfo {volume}
  {757}},\ \bibinfo {pages} {479} (\bibinfo {year} {2005})},\ \Eprint
  {http://arxiv.org/abs/nucl-th/0403039} {arXiv:nucl-th/0403039} \BibitemShut
  {NoStop}%
\bibitem [{\citenamefont {Schaefer}\ \emph {et~al.}(2007)\citenamefont
  {Schaefer}, \citenamefont {Pawlowski},\ and\ \citenamefont
  {Wambach}}]{Schaefer:2007pw}%
  \BibitemOpen
  \bibfield  {author} {\bibinfo {author} {\bibfnamefont {B.-J.}\ \bibnamefont
  {Schaefer}}, \bibinfo {author} {\bibfnamefont {J.~M.}\ \bibnamefont
  {Pawlowski}}, \ and\ \bibinfo {author} {\bibfnamefont {J.}~\bibnamefont
  {Wambach}},\ }\href {\doibase 10.1103/PhysRevD.76.074023} {\bibfield
  {journal} {\bibinfo  {journal} {Phys. Rev. D}\ }\textbf {\bibinfo {volume}
  {76}},\ \bibinfo {pages} {074023} (\bibinfo {year} {2007})},\ \Eprint
  {http://arxiv.org/abs/0704.3234} {arXiv:0704.3234 [hep-ph]} \BibitemShut
  {NoStop}%
\bibitem [{\citenamefont {Mao}\ \emph {et~al.}(2010)\citenamefont {Mao},
  \citenamefont {Jin},\ and\ \citenamefont {Huang}}]{Mao:2009aq}%
  \BibitemOpen
  \bibfield  {author} {\bibinfo {author} {\bibfnamefont {H.}~\bibnamefont
  {Mao}}, \bibinfo {author} {\bibfnamefont {J.}~\bibnamefont {Jin}}, \ and\
  \bibinfo {author} {\bibfnamefont {M.}~\bibnamefont {Huang}},\ }\href
  {\doibase 10.1088/0954-3899/37/3/035001} {\bibfield  {journal} {\bibinfo
  {journal} {J. Phys. G}\ }\textbf {\bibinfo {volume} {37}},\ \bibinfo {pages}
  {035001} (\bibinfo {year} {2010})},\ \Eprint {http://arxiv.org/abs/0906.1324}
  {arXiv:0906.1324 [hep-ph]} \BibitemShut {NoStop}%
\bibitem [{\citenamefont {Schaefer}\ and\ \citenamefont
  {Wagner}(2012{\natexlab{a}})}]{Schaefer:2011ex}%
  \BibitemOpen
  \bibfield  {author} {\bibinfo {author} {\bibfnamefont {B.~J.}\ \bibnamefont
  {Schaefer}}\ and\ \bibinfo {author} {\bibfnamefont {M.}~\bibnamefont
  {Wagner}},\ }\href {\doibase 10.1103/PhysRevD.85.034027} {\bibfield
  {journal} {\bibinfo  {journal} {Phys. Rev. D}\ }\textbf {\bibinfo {volume}
  {85}},\ \bibinfo {pages} {034027} (\bibinfo {year} {2012}{\natexlab{a}})},\
  \Eprint {http://arxiv.org/abs/1111.6871} {arXiv:1111.6871 [hep-ph]}
  \BibitemShut {NoStop}%
\bibitem [{\citenamefont {Schaefer}\ and\ \citenamefont
  {Wagner}(2012{\natexlab{b}})}]{Schaefer:2012gy}%
  \BibitemOpen
  \bibfield  {author} {\bibinfo {author} {\bibfnamefont {B.-J.}\ \bibnamefont
  {Schaefer}}\ and\ \bibinfo {author} {\bibfnamefont {M.}~\bibnamefont
  {Wagner}},\ }\href {\doibase 10.2478/s11534-012-0115-y} {\bibfield  {journal}
  {\bibinfo  {journal} {Central Eur. J. Phys.}\ }\textbf {\bibinfo {volume}
  {10}},\ \bibinfo {pages} {1326} (\bibinfo {year} {2012}{\natexlab{b}})},\
  \Eprint {http://arxiv.org/abs/1203.1883} {arXiv:1203.1883 [hep-ph]}
  \BibitemShut {NoStop}%
\bibitem [{\citenamefont {Qin}\ \emph {et~al.}(2011)\citenamefont {Qin},
  \citenamefont {Chang}, \citenamefont {Chen}, \citenamefont {Liu},\ and\
  \citenamefont {Roberts}}]{Qin:2010nq}%
  \BibitemOpen
  \bibfield  {author} {\bibinfo {author} {\bibfnamefont {S.-x.}\ \bibnamefont
  {Qin}}, \bibinfo {author} {\bibfnamefont {L.}~\bibnamefont {Chang}}, \bibinfo
  {author} {\bibfnamefont {H.}~\bibnamefont {Chen}}, \bibinfo {author}
  {\bibfnamefont {Y.-x.}\ \bibnamefont {Liu}}, \ and\ \bibinfo {author}
  {\bibfnamefont {C.~D.}\ \bibnamefont {Roberts}},\ }\href {\doibase
  10.1103/PhysRevLett.106.172301} {\bibfield  {journal} {\bibinfo  {journal}
  {Phys. Rev. Lett.}\ }\textbf {\bibinfo {volume} {106}},\ \bibinfo {pages}
  {172301} (\bibinfo {year} {2011})},\ \Eprint {http://arxiv.org/abs/1011.2876}
  {arXiv:1011.2876 [nucl-th]} \BibitemShut {NoStop}%
\bibitem [{\citenamefont {Sun}\ \emph {et~al.}(2023{\natexlab{b}})\citenamefont
  {Sun}, \citenamefont {Xu},\ and\ \citenamefont {Huang}}]{Sun:2023kuu}%
  \BibitemOpen
  \bibfield  {author} {\bibinfo {author} {\bibfnamefont {F.}~\bibnamefont
  {Sun}}, \bibinfo {author} {\bibfnamefont {K.}~\bibnamefont {Xu}}, \ and\
  \bibinfo {author} {\bibfnamefont {M.}~\bibnamefont {Huang}},\ }\href
  {\doibase 10.1103/PhysRevD.108.096007} {\bibfield  {journal} {\bibinfo
  {journal} {Phys. Rev. D}\ }\textbf {\bibinfo {volume} {108}},\ \bibinfo
  {pages} {096007} (\bibinfo {year} {2023}{\natexlab{b}})},\ \Eprint
  {http://arxiv.org/abs/2307.14402} {arXiv:2307.14402 [hep-ph]} \BibitemShut
  {NoStop}%
\bibitem [{\citenamefont {Cao}(2021)}]{Cao:2021rwx}%
  \BibitemOpen
  \bibfield  {author} {\bibinfo {author} {\bibfnamefont {G.}~\bibnamefont
  {Cao}},\ }\href {\doibase 10.1140/epja/s10050-021-00570-0} {\bibfield
  {journal} {\bibinfo  {journal} {Eur. Phys. J. A}\ }\textbf {\bibinfo {volume}
  {57}},\ \bibinfo {pages} {264} (\bibinfo {year} {2021})},\ \Eprint
  {http://arxiv.org/abs/2103.00456} {arXiv:2103.00456 [hep-ph]} \BibitemShut
  {NoStop}%
\bibitem [{\citenamefont {Sun}\ \emph {et~al.}(2024)\citenamefont {Sun},
  \citenamefont {Shao}, \citenamefont {Wen}, \citenamefont {Xu},\ and\
  \citenamefont {Huang}}]{Sun:2024anu}%
  \BibitemOpen
  \bibfield  {author} {\bibinfo {author} {\bibfnamefont {F.}~\bibnamefont
  {Sun}}, \bibinfo {author} {\bibfnamefont {J.}~\bibnamefont {Shao}}, \bibinfo
  {author} {\bibfnamefont {R.}~\bibnamefont {Wen}}, \bibinfo {author}
  {\bibfnamefont {K.}~\bibnamefont {Xu}}, \ and\ \bibinfo {author}
  {\bibfnamefont {M.}~\bibnamefont {Huang}},\ }\href {\doibase
  10.1103/PhysRevD.109.116017} {\bibfield  {journal} {\bibinfo  {journal}
  {Phys. Rev. D}\ }\textbf {\bibinfo {volume} {109}},\ \bibinfo {pages}
  {116017} (\bibinfo {year} {2024})},\ \Eprint
  {http://arxiv.org/abs/2402.16595} {arXiv:2402.16595 [hep-ph]} \BibitemShut
  {NoStop}%
\bibitem [{\citenamefont {Luecker}\ \emph {et~al.}(2013)\citenamefont
  {Luecker}, \citenamefont {Fischer}, \citenamefont {Fister},\ and\
  \citenamefont {Pawlowski}}]{Luecker:2013oda}%
  \BibitemOpen
  \bibfield  {author} {\bibinfo {author} {\bibfnamefont {J.}~\bibnamefont
  {Luecker}}, \bibinfo {author} {\bibfnamefont {C.~S.}\ \bibnamefont
  {Fischer}}, \bibinfo {author} {\bibfnamefont {L.}~\bibnamefont {Fister}}, \
  and\ \bibinfo {author} {\bibfnamefont {J.~M.}\ \bibnamefont {Pawlowski}},\
  }\href {\doibase 10.22323/1.185.0057} {\bibfield  {journal} {\bibinfo
  {journal} {PoS}\ }\textbf {\bibinfo {volume} {CPOD2013}},\ \bibinfo {pages}
  {057} (\bibinfo {year} {2013})},\ \Eprint {http://arxiv.org/abs/1308.4509}
  {arXiv:1308.4509 [hep-ph]} \BibitemShut {NoStop}%
\bibitem [{\citenamefont {Fu}\ \emph {et~al.}(2016)\citenamefont {Fu},
  \citenamefont {Pawlowski}, \citenamefont {Rennecke},\ and\ \citenamefont
  {Schaefer}}]{Fu:2016tey}%
  \BibitemOpen
  \bibfield  {author} {\bibinfo {author} {\bibfnamefont {W.-j.}\ \bibnamefont
  {Fu}}, \bibinfo {author} {\bibfnamefont {J.~M.}\ \bibnamefont {Pawlowski}},
  \bibinfo {author} {\bibfnamefont {F.}~\bibnamefont {Rennecke}}, \ and\
  \bibinfo {author} {\bibfnamefont {B.-J.}\ \bibnamefont {Schaefer}},\ }\href
  {\doibase 10.1103/PhysRevD.94.116020} {\bibfield  {journal} {\bibinfo
  {journal} {Phys. Rev. D}\ }\textbf {\bibinfo {volume} {94}},\ \bibinfo
  {pages} {116020} (\bibinfo {year} {2016})},\ \Eprint
  {http://arxiv.org/abs/1608.04302} {arXiv:1608.04302 [hep-ph]} \BibitemShut
  {NoStop}%
\bibitem [{\citenamefont {Chen}\ \emph
  {et~al.}(2021{\natexlab{a}})\citenamefont {Chen}, \citenamefont {Wen},\ and\
  \citenamefont {Fu}}]{Chen:2021iuo}%
  \BibitemOpen
  \bibfield  {author} {\bibinfo {author} {\bibfnamefont {Y.-r.}\ \bibnamefont
  {Chen}}, \bibinfo {author} {\bibfnamefont {R.}~\bibnamefont {Wen}}, \ and\
  \bibinfo {author} {\bibfnamefont {W.-j.}\ \bibnamefont {Fu}},\ }\href
  {\doibase 10.1103/PhysRevD.104.054009} {\bibfield  {journal} {\bibinfo
  {journal} {Phys. Rev. D}\ }\textbf {\bibinfo {volume} {104}},\ \bibinfo
  {pages} {054009} (\bibinfo {year} {2021}{\natexlab{a}})},\ \Eprint
  {http://arxiv.org/abs/2101.08484} {arXiv:2101.08484 [hep-ph]} \BibitemShut
  {NoStop}%
\bibitem [{\citenamefont {Braun}\ \emph {et~al.}(2023)\citenamefont {Braun}
  \emph {et~al.}}]{Braun:2023qak}%
  \BibitemOpen
  \bibfield  {author} {\bibinfo {author} {\bibfnamefont {J.}~\bibnamefont
  {Braun}} \emph {et~al.},\ }\href@noop {} {\  (\bibinfo {year} {2023})},\
  \Eprint {http://arxiv.org/abs/2310.19853} {arXiv:2310.19853 [hep-ph]}
  \BibitemShut {NoStop}%
\bibitem [{\citenamefont {Maldacena}(1998)}]{Maldacena:1997re}%
  \BibitemOpen
  \bibfield  {author} {\bibinfo {author} {\bibfnamefont {J.~M.}\ \bibnamefont
  {Maldacena}},\ }\href {\doibase 10.4310/ATMP.1998.v2.n2.a1} {\bibfield
  {journal} {\bibinfo  {journal} {Adv. Theor. Math. Phys.}\ }\textbf {\bibinfo
  {volume} {2}},\ \bibinfo {pages} {231} (\bibinfo {year} {1998})},\ \Eprint
  {http://arxiv.org/abs/hep-th/9711200} {arXiv:hep-th/9711200} \BibitemShut
  {NoStop}%
\bibitem [{\citenamefont {Gubser}\ \emph {et~al.}(1998)\citenamefont {Gubser},
  \citenamefont {Klebanov},\ and\ \citenamefont {Polyakov}}]{Gubser:1998bc}%
  \BibitemOpen
  \bibfield  {author} {\bibinfo {author} {\bibfnamefont {S.~S.}\ \bibnamefont
  {Gubser}}, \bibinfo {author} {\bibfnamefont {I.~R.}\ \bibnamefont
  {Klebanov}}, \ and\ \bibinfo {author} {\bibfnamefont {A.~M.}\ \bibnamefont
  {Polyakov}},\ }\href {\doibase 10.1016/S0370-2693(98)00377-3} {\bibfield
  {journal} {\bibinfo  {journal} {Phys. Lett. B}\ }\textbf {\bibinfo {volume}
  {428}},\ \bibinfo {pages} {105} (\bibinfo {year} {1998})},\ \Eprint
  {http://arxiv.org/abs/hep-th/9802109} {arXiv:hep-th/9802109} \BibitemShut
  {NoStop}%
\bibitem [{\citenamefont {Witten}(1998)}]{Witten:1998qj}%
  \BibitemOpen
  \bibfield  {author} {\bibinfo {author} {\bibfnamefont {E.}~\bibnamefont
  {Witten}},\ }\href {\doibase 10.4310/ATMP.1998.v2.n2.a2} {\bibfield
  {journal} {\bibinfo  {journal} {Adv. Theor. Math. Phys.}\ }\textbf {\bibinfo
  {volume} {2}},\ \bibinfo {pages} {253} (\bibinfo {year} {1998})},\ \Eprint
  {http://arxiv.org/abs/hep-th/9802150} {arXiv:hep-th/9802150} \BibitemShut
  {NoStop}%
\bibitem [{\citenamefont {DeWolfe}\ \emph
  {et~al.}(2011{\natexlab{a}})\citenamefont {DeWolfe}, \citenamefont {Gubser},\
  and\ \citenamefont {Rosen}}]{DeWolfe:2010he}%
  \BibitemOpen
  \bibfield  {author} {\bibinfo {author} {\bibfnamefont {O.}~\bibnamefont
  {DeWolfe}}, \bibinfo {author} {\bibfnamefont {S.~S.}\ \bibnamefont {Gubser}},
  \ and\ \bibinfo {author} {\bibfnamefont {C.}~\bibnamefont {Rosen}},\ }\href
  {\doibase 10.1103/PhysRevD.83.086005} {\bibfield  {journal} {\bibinfo
  {journal} {Phys. Rev. D}\ }\textbf {\bibinfo {volume} {83}},\ \bibinfo
  {pages} {086005} (\bibinfo {year} {2011}{\natexlab{a}})},\ \Eprint
  {http://arxiv.org/abs/1012.1864} {arXiv:1012.1864 [hep-th]} \BibitemShut
  {NoStop}%
\bibitem [{\citenamefont {DeWolfe}\ \emph
  {et~al.}(2011{\natexlab{b}})\citenamefont {DeWolfe}, \citenamefont {Gubser},\
  and\ \citenamefont {Rosen}}]{DeWolfe:2011ts}%
  \BibitemOpen
  \bibfield  {author} {\bibinfo {author} {\bibfnamefont {O.}~\bibnamefont
  {DeWolfe}}, \bibinfo {author} {\bibfnamefont {S.~S.}\ \bibnamefont {Gubser}},
  \ and\ \bibinfo {author} {\bibfnamefont {C.}~\bibnamefont {Rosen}},\ }\href
  {\doibase 10.1103/PhysRevD.84.126014} {\bibfield  {journal} {\bibinfo
  {journal} {Phys. Rev. D}\ }\textbf {\bibinfo {volume} {84}},\ \bibinfo
  {pages} {126014} (\bibinfo {year} {2011}{\natexlab{b}})},\ \Eprint
  {http://arxiv.org/abs/1108.2029} {arXiv:1108.2029 [hep-th]} \BibitemShut
  {NoStop}%
\bibitem [{\citenamefont {Chelabi}\ \emph
  {et~al.}(2016{\natexlab{a}})\citenamefont {Chelabi}, \citenamefont {Fang},
  \citenamefont {Huang}, \citenamefont {Li},\ and\ \citenamefont
  {Wu}}]{Chelabi:2015cwn}%
  \BibitemOpen
  \bibfield  {author} {\bibinfo {author} {\bibfnamefont {K.}~\bibnamefont
  {Chelabi}}, \bibinfo {author} {\bibfnamefont {Z.}~\bibnamefont {Fang}},
  \bibinfo {author} {\bibfnamefont {M.}~\bibnamefont {Huang}}, \bibinfo
  {author} {\bibfnamefont {D.}~\bibnamefont {Li}}, \ and\ \bibinfo {author}
  {\bibfnamefont {Y.-L.}\ \bibnamefont {Wu}},\ }\href {\doibase
  10.1103/PhysRevD.93.101901} {\bibfield  {journal} {\bibinfo  {journal} {Phys.
  Rev. D}\ }\textbf {\bibinfo {volume} {93}},\ \bibinfo {pages} {101901}
  (\bibinfo {year} {2016}{\natexlab{a}})},\ \Eprint
  {http://arxiv.org/abs/1511.02721} {arXiv:1511.02721 [hep-ph]} \BibitemShut
  {NoStop}%
\bibitem [{\citenamefont {Chelabi}\ \emph
  {et~al.}(2016{\natexlab{b}})\citenamefont {Chelabi}, \citenamefont {Fang},
  \citenamefont {Huang}, \citenamefont {Li},\ and\ \citenamefont
  {Wu}}]{Chelabi:2015gpc}%
  \BibitemOpen
  \bibfield  {author} {\bibinfo {author} {\bibfnamefont {K.}~\bibnamefont
  {Chelabi}}, \bibinfo {author} {\bibfnamefont {Z.}~\bibnamefont {Fang}},
  \bibinfo {author} {\bibfnamefont {M.}~\bibnamefont {Huang}}, \bibinfo
  {author} {\bibfnamefont {D.}~\bibnamefont {Li}}, \ and\ \bibinfo {author}
  {\bibfnamefont {Y.-L.}\ \bibnamefont {Wu}},\ }\href {\doibase
  10.1007/JHEP04(2016)036} {\bibfield  {journal} {\bibinfo  {journal} {JHEP}\
  }\textbf {\bibinfo {volume} {04}},\ \bibinfo {pages} {036} (\bibinfo {year}
  {2016}{\natexlab{b}})},\ \Eprint {http://arxiv.org/abs/1512.06493}
  {arXiv:1512.06493 [hep-ph]} \BibitemShut {NoStop}%
\bibitem [{\citenamefont {Li}\ and\ \citenamefont {Huang}(2017)}]{Li:2016smq}%
  \BibitemOpen
  \bibfield  {author} {\bibinfo {author} {\bibfnamefont {D.}~\bibnamefont
  {Li}}\ and\ \bibinfo {author} {\bibfnamefont {M.}~\bibnamefont {Huang}},\
  }\href {\doibase 10.1007/JHEP02(2017)042} {\bibfield  {journal} {\bibinfo
  {journal} {JHEP}\ }\textbf {\bibinfo {volume} {02}},\ \bibinfo {pages} {042}
  (\bibinfo {year} {2017})},\ \Eprint {http://arxiv.org/abs/1610.09814}
  {arXiv:1610.09814 [hep-ph]} \BibitemShut {NoStop}%
\bibitem [{\citenamefont {Chen}\ \emph
  {et~al.}(2019{\natexlab{a}})\citenamefont {Chen}, \citenamefont {He},
  \citenamefont {Huang},\ and\ \citenamefont {Li}}]{Chen:2018msc}%
  \BibitemOpen
  \bibfield  {author} {\bibinfo {author} {\bibfnamefont {J.}~\bibnamefont
  {Chen}}, \bibinfo {author} {\bibfnamefont {S.}~\bibnamefont {He}}, \bibinfo
  {author} {\bibfnamefont {M.}~\bibnamefont {Huang}}, \ and\ \bibinfo {author}
  {\bibfnamefont {D.}~\bibnamefont {Li}},\ }\href {\doibase
  10.1007/JHEP01(2019)165} {\bibfield  {journal} {\bibinfo  {journal} {JHEP}\
  }\textbf {\bibinfo {volume} {01}},\ \bibinfo {pages} {165} (\bibinfo {year}
  {2019}{\natexlab{a}})},\ \Eprint {http://arxiv.org/abs/1810.07019}
  {arXiv:1810.07019 [hep-ph]} \BibitemShut {NoStop}%
\bibitem [{\citenamefont {Chen}\ \emph {et~al.}(2020)\citenamefont {Chen},
  \citenamefont {Li}, \citenamefont {Hou},\ and\ \citenamefont
  {Huang}}]{Chen:2019rez}%
  \BibitemOpen
  \bibfield  {author} {\bibinfo {author} {\bibfnamefont {X.}~\bibnamefont
  {Chen}}, \bibinfo {author} {\bibfnamefont {D.}~\bibnamefont {Li}}, \bibinfo
  {author} {\bibfnamefont {D.}~\bibnamefont {Hou}}, \ and\ \bibinfo {author}
  {\bibfnamefont {M.}~\bibnamefont {Huang}},\ }\href {\doibase
  10.1007/JHEP03(2020)073} {\bibfield  {journal} {\bibinfo  {journal} {JHEP}\
  }\textbf {\bibinfo {volume} {03}},\ \bibinfo {pages} {073} (\bibinfo {year}
  {2020})},\ \Eprint {http://arxiv.org/abs/1908.02000} {arXiv:1908.02000
  [hep-ph]} \BibitemShut {NoStop}%
\bibitem [{\citenamefont {Choun}\ and\ \citenamefont
  {Sin}(2020)}]{Choun:2019xyo}%
  \BibitemOpen
  \bibfield  {author} {\bibinfo {author} {\bibfnamefont {Y.-S.}\ \bibnamefont
  {Choun}}\ and\ \bibinfo {author} {\bibfnamefont {S.-J.}\ \bibnamefont
  {Sin}},\ }\href {\doibase 10.1016/j.physletb.2020.135433} {\bibfield
  {journal} {\bibinfo  {journal} {Phys. Lett. B}\ }\textbf {\bibinfo {volume}
  {805}},\ \bibinfo {pages} {135433} (\bibinfo {year} {2020})},\ \Eprint
  {http://arxiv.org/abs/1910.02383} {arXiv:1910.02383 [hep-th]} \BibitemShut
  {NoStop}%
\bibitem [{\citenamefont {Mamani}\ \emph {et~al.}(2020)\citenamefont {Mamani},
  \citenamefont {Flores},\ and\ \citenamefont {Zanchin}}]{Mamani:2020pks}%
  \BibitemOpen
  \bibfield  {author} {\bibinfo {author} {\bibfnamefont {L.~A.~H.}\
  \bibnamefont {Mamani}}, \bibinfo {author} {\bibfnamefont {C.~V.}\
  \bibnamefont {Flores}}, \ and\ \bibinfo {author} {\bibfnamefont {V.~T.}\
  \bibnamefont {Zanchin}},\ }\href {\doibase 10.1103/PhysRevD.102.066006}
  {\bibfield  {journal} {\bibinfo  {journal} {Phys. Rev. D}\ }\textbf {\bibinfo
  {volume} {102}},\ \bibinfo {pages} {066006} (\bibinfo {year} {2020})},\
  \Eprint {http://arxiv.org/abs/2006.09401} {arXiv:2006.09401 [hep-th]}
  \BibitemShut {NoStop}%
\bibitem [{\citenamefont {Chen}\ \emph
  {et~al.}(2024{\natexlab{a}})\citenamefont {Chen}, \citenamefont {Chen},
  \citenamefont {Li},\ and\ \citenamefont {Huang}}]{Chen:2024jet}%
  \BibitemOpen
  \bibfield  {author} {\bibinfo {author} {\bibfnamefont {Y.}~\bibnamefont
  {Chen}}, \bibinfo {author} {\bibfnamefont {X.}~\bibnamefont {Chen}}, \bibinfo
  {author} {\bibfnamefont {D.}~\bibnamefont {Li}}, \ and\ \bibinfo {author}
  {\bibfnamefont {M.}~\bibnamefont {Huang}},\ }\href@noop {} {\  (\bibinfo
  {year} {2024}{\natexlab{a}})},\ \Eprint {http://arxiv.org/abs/2405.06386}
  {arXiv:2405.06386 [hep-ph]} \BibitemShut {NoStop}%
\bibitem [{\citenamefont {Ahmed}\ \emph {et~al.}(2024)\citenamefont {Ahmed},
  \citenamefont {Kawaguchi},\ and\ \citenamefont {Huang}}]{Ahmed:2024rbj}%
  \BibitemOpen
  \bibfield  {author} {\bibinfo {author} {\bibfnamefont {H.~A.}\ \bibnamefont
  {Ahmed}}, \bibinfo {author} {\bibfnamefont {M.}~\bibnamefont {Kawaguchi}}, \
  and\ \bibinfo {author} {\bibfnamefont {M.}~\bibnamefont {Huang}},\ }\href
  {\doibase 10.1103/PhysRevD.110.046002} {\bibfield  {journal} {\bibinfo
  {journal} {Phys. Rev. D}\ }\textbf {\bibinfo {volume} {110}},\ \bibinfo
  {pages} {046002} (\bibinfo {year} {2024})},\ \Eprint
  {http://arxiv.org/abs/2401.04355} {arXiv:2401.04355 [hep-ph]} \BibitemShut
  {NoStop}%
\bibitem [{\citenamefont {Wang}\ and\ \citenamefont
  {Feng}(2024)}]{Wang:2024szr}%
  \BibitemOpen
  \bibfield  {author} {\bibinfo {author} {\bibfnamefont {J.-H.}\ \bibnamefont
  {Wang}}\ and\ \bibinfo {author} {\bibfnamefont {S.-Q.}\ \bibnamefont
  {Feng}},\ }\href {\doibase 10.1103/PhysRevD.109.066019} {\bibfield  {journal}
  {\bibinfo  {journal} {Phys. Rev. D}\ }\textbf {\bibinfo {volume} {109}},\
  \bibinfo {pages} {066019} (\bibinfo {year} {2024})},\ \Eprint
  {http://arxiv.org/abs/2403.01814} {arXiv:2403.01814 [hep-ph]} \BibitemShut
  {NoStop}%
\bibitem [{\citenamefont {Li}(2024)}]{Li:2024lrh}%
  \BibitemOpen
  \bibfield  {author} {\bibinfo {author} {\bibfnamefont {Z.}~\bibnamefont
  {Li}},\ }\href {\doibase 10.1103/PhysRevD.110.046012} {\bibfield  {journal}
  {\bibinfo  {journal} {Phys. Rev. D}\ }\textbf {\bibinfo {volume} {110}},\
  \bibinfo {pages} {046012} (\bibinfo {year} {2024})},\ \Eprint
  {http://arxiv.org/abs/2402.02944} {arXiv:2402.02944 [hep-th]} \BibitemShut
  {NoStop}%
\bibitem [{\citenamefont {Zhao}\ \emph {et~al.}(2024)\citenamefont {Zhao},
  \citenamefont {He}, \citenamefont {Hou}, \citenamefont {Li},\ and\
  \citenamefont {Li}}]{Zhao:2023gur}%
  \BibitemOpen
  \bibfield  {author} {\bibinfo {author} {\bibfnamefont {Y.-Q.}\ \bibnamefont
  {Zhao}}, \bibinfo {author} {\bibfnamefont {S.}~\bibnamefont {He}}, \bibinfo
  {author} {\bibfnamefont {D.}~\bibnamefont {Hou}}, \bibinfo {author}
  {\bibfnamefont {L.}~\bibnamefont {Li}}, \ and\ \bibinfo {author}
  {\bibfnamefont {Z.}~\bibnamefont {Li}},\ }\href {\doibase
  10.1103/PhysRevD.109.086015} {\bibfield  {journal} {\bibinfo  {journal}
  {Phys. Rev. D}\ }\textbf {\bibinfo {volume} {109}},\ \bibinfo {pages}
  {086015} (\bibinfo {year} {2024})},\ \Eprint
  {http://arxiv.org/abs/2310.13432} {arXiv:2310.13432 [hep-ph]} \BibitemShut
  {NoStop}%
\bibitem [{\citenamefont {Rougemont}\ \emph {et~al.}(2023)\citenamefont
  {Rougemont}, \citenamefont {Grefa}, \citenamefont {Hippert}, \citenamefont
  {Noronha}, \citenamefont {Noronha-Hostler}, \citenamefont {Portillo},\ and\
  \citenamefont {Ratti}}]{Rougemont:2023gfz}%
  \BibitemOpen
  \bibfield  {author} {\bibinfo {author} {\bibfnamefont {R.}~\bibnamefont
  {Rougemont}}, \bibinfo {author} {\bibfnamefont {J.}~\bibnamefont {Grefa}},
  \bibinfo {author} {\bibfnamefont {M.}~\bibnamefont {Hippert}}, \bibinfo
  {author} {\bibfnamefont {J.}~\bibnamefont {Noronha}}, \bibinfo {author}
  {\bibfnamefont {J.}~\bibnamefont {Noronha-Hostler}}, \bibinfo {author}
  {\bibfnamefont {I.}~\bibnamefont {Portillo}}, \ and\ \bibinfo {author}
  {\bibfnamefont {C.}~\bibnamefont {Ratti}},\ }\href@noop {} {\  (\bibinfo
  {year} {2023})},\ \Eprint {http://arxiv.org/abs/2307.03885} {arXiv:2307.03885
  [nucl-th]} \BibitemShut {NoStop}%
\bibitem [{\citenamefont {Zhao}\ \emph {et~al.}(2023)\citenamefont {Zhao},
  \citenamefont {He}, \citenamefont {Hou}, \citenamefont {Li},\ and\
  \citenamefont {Li}}]{Zhao:2022uxc}%
  \BibitemOpen
  \bibfield  {author} {\bibinfo {author} {\bibfnamefont {Y.-Q.}\ \bibnamefont
  {Zhao}}, \bibinfo {author} {\bibfnamefont {S.}~\bibnamefont {He}}, \bibinfo
  {author} {\bibfnamefont {D.}~\bibnamefont {Hou}}, \bibinfo {author}
  {\bibfnamefont {L.}~\bibnamefont {Li}}, \ and\ \bibinfo {author}
  {\bibfnamefont {Z.}~\bibnamefont {Li}},\ }\href {\doibase
  10.1007/JHEP04(2023)115} {\bibfield  {journal} {\bibinfo  {journal} {JHEP}\
  }\textbf {\bibinfo {volume} {04}},\ \bibinfo {pages} {115} (\bibinfo {year}
  {2023})},\ \Eprint {http://arxiv.org/abs/2212.14662} {arXiv:2212.14662
  [hep-ph]} \BibitemShut {NoStop}%
\bibitem [{\citenamefont {Yang}\ and\ \citenamefont
  {Yuan}(2022)}]{Yang:2020hun}%
  \BibitemOpen
  \bibfield  {author} {\bibinfo {author} {\bibfnamefont {Y.}~\bibnamefont
  {Yang}}\ and\ \bibinfo {author} {\bibfnamefont {P.-H.}\ \bibnamefont
  {Yuan}},\ }\href {\doibase 10.1016/j.physletb.2022.137212} {\bibfield
  {journal} {\bibinfo  {journal} {Phys. Lett. B}\ }\textbf {\bibinfo {volume}
  {832}},\ \bibinfo {pages} {137212} (\bibinfo {year} {2022})},\ \Eprint
  {http://arxiv.org/abs/2011.11941} {arXiv:2011.11941 [hep-th]} \BibitemShut
  {NoStop}%
\bibitem [{\citenamefont {Chen}\ \emph
  {et~al.}(2019{\natexlab{b}})\citenamefont {Chen}, \citenamefont {Li},\ and\
  \citenamefont {Huang}}]{Chen:2018vty}%
  \BibitemOpen
  \bibfield  {author} {\bibinfo {author} {\bibfnamefont {X.}~\bibnamefont
  {Chen}}, \bibinfo {author} {\bibfnamefont {D.}~\bibnamefont {Li}}, \ and\
  \bibinfo {author} {\bibfnamefont {M.}~\bibnamefont {Huang}},\ }\href
  {\doibase 10.1088/1674-1137/43/2/023105} {\bibfield  {journal} {\bibinfo
  {journal} {Chin. Phys. C}\ }\textbf {\bibinfo {volume} {43}},\ \bibinfo
  {pages} {023105} (\bibinfo {year} {2019}{\natexlab{b}})},\ \Eprint
  {http://arxiv.org/abs/1810.02136} {arXiv:1810.02136 [hep-ph]} \BibitemShut
  {NoStop}%
\bibitem [{\citenamefont {Fu}\ \emph {et~al.}(2024)\citenamefont {Fu},
  \citenamefont {He}, \citenamefont {Li},\ and\ \citenamefont
  {Li}}]{Fu:2024wkn}%
  \BibitemOpen
  \bibfield  {author} {\bibinfo {author} {\bibfnamefont {Q.}~\bibnamefont
  {Fu}}, \bibinfo {author} {\bibfnamefont {S.}~\bibnamefont {He}}, \bibinfo
  {author} {\bibfnamefont {L.}~\bibnamefont {Li}}, \ and\ \bibinfo {author}
  {\bibfnamefont {Z.}~\bibnamefont {Li}},\ }\href@noop {} {\  (\bibinfo {year}
  {2024})},\ \Eprint {http://arxiv.org/abs/2404.12109} {arXiv:2404.12109
  [hep-ph]} \BibitemShut {NoStop}%
\bibitem [{\citenamefont {Cai}\ \emph {et~al.}(2024)\citenamefont {Cai},
  \citenamefont {He}, \citenamefont {Li},\ and\ \citenamefont
  {Zeng}}]{Cai:2024eqa}%
  \BibitemOpen
  \bibfield  {author} {\bibinfo {author} {\bibfnamefont {R.-G.}\ \bibnamefont
  {Cai}}, \bibinfo {author} {\bibfnamefont {S.}~\bibnamefont {He}}, \bibinfo
  {author} {\bibfnamefont {L.}~\bibnamefont {Li}}, \ and\ \bibinfo {author}
  {\bibfnamefont {H.-A.}\ \bibnamefont {Zeng}},\ }\href@noop {} {\  (\bibinfo
  {year} {2024})},\ \Eprint {http://arxiv.org/abs/2406.12772} {arXiv:2406.12772
  [hep-th]} \BibitemShut {NoStop}%
\bibitem [{\citenamefont {He}\ \emph {et~al.}(2013)\citenamefont {He},
  \citenamefont {Wu}, \citenamefont {Yang},\ and\ \citenamefont
  {Yuan}}]{He:2013qq}%
  \BibitemOpen
  \bibfield  {author} {\bibinfo {author} {\bibfnamefont {S.}~\bibnamefont
  {He}}, \bibinfo {author} {\bibfnamefont {S.-Y.}\ \bibnamefont {Wu}}, \bibinfo
  {author} {\bibfnamefont {Y.}~\bibnamefont {Yang}}, \ and\ \bibinfo {author}
  {\bibfnamefont {P.-H.}\ \bibnamefont {Yuan}},\ }\href {\doibase
  10.1007/JHEP04(2013)093} {\bibfield  {journal} {\bibinfo  {journal} {JHEP}\
  }\textbf {\bibinfo {volume} {04}},\ \bibinfo {pages} {093} (\bibinfo {year}
  {2013})},\ \Eprint {http://arxiv.org/abs/1301.0385} {arXiv:1301.0385
  [hep-th]} \BibitemShut {NoStop}%
\bibitem [{\citenamefont {Yang}\ and\ \citenamefont
  {Yuan}(2014)}]{Yang:2014bqa}%
  \BibitemOpen
  \bibfield  {author} {\bibinfo {author} {\bibfnamefont {Y.}~\bibnamefont
  {Yang}}\ and\ \bibinfo {author} {\bibfnamefont {P.-H.}\ \bibnamefont
  {Yuan}},\ }\href {\doibase 10.1007/JHEP11(2014)149} {\bibfield  {journal}
  {\bibinfo  {journal} {JHEP}\ }\textbf {\bibinfo {volume} {11}},\ \bibinfo
  {pages} {149} (\bibinfo {year} {2014})},\ \Eprint
  {http://arxiv.org/abs/1406.1865} {arXiv:1406.1865 [hep-th]} \BibitemShut
  {NoStop}%
\bibitem [{\citenamefont {Yang}\ and\ \citenamefont
  {Yuan}(2015)}]{Yang:2015aia}%
  \BibitemOpen
  \bibfield  {author} {\bibinfo {author} {\bibfnamefont {Y.}~\bibnamefont
  {Yang}}\ and\ \bibinfo {author} {\bibfnamefont {P.-H.}\ \bibnamefont
  {Yuan}},\ }\href {\doibase 10.1007/JHEP12(2015)161} {\bibfield  {journal}
  {\bibinfo  {journal} {JHEP}\ }\textbf {\bibinfo {volume} {12}},\ \bibinfo
  {pages} {161} (\bibinfo {year} {2015})},\ \Eprint
  {http://arxiv.org/abs/1506.05930} {arXiv:1506.05930 [hep-th]} \BibitemShut
  {NoStop}%
\bibitem [{\citenamefont {Dudal}\ and\ \citenamefont
  {Mahapatra}(2017)}]{Dudal:2017max}%
  \BibitemOpen
  \bibfield  {author} {\bibinfo {author} {\bibfnamefont {D.}~\bibnamefont
  {Dudal}}\ and\ \bibinfo {author} {\bibfnamefont {S.}~\bibnamefont
  {Mahapatra}},\ }\href {\doibase 10.1103/PhysRevD.96.126010} {\bibfield
  {journal} {\bibinfo  {journal} {Phys. Rev. D}\ }\textbf {\bibinfo {volume}
  {96}},\ \bibinfo {pages} {126010} (\bibinfo {year} {2017})},\ \Eprint
  {http://arxiv.org/abs/1708.06995} {arXiv:1708.06995 [hep-th]} \BibitemShut
  {NoStop}%
\bibitem [{\citenamefont {Dudal}\ and\ \citenamefont
  {Mahapatra}(2018)}]{Dudal:2018ztm}%
  \BibitemOpen
  \bibfield  {author} {\bibinfo {author} {\bibfnamefont {D.}~\bibnamefont
  {Dudal}}\ and\ \bibinfo {author} {\bibfnamefont {S.}~\bibnamefont
  {Mahapatra}},\ }\href {\doibase 10.1007/JHEP07(2018)120} {\bibfield
  {journal} {\bibinfo  {journal} {JHEP}\ }\textbf {\bibinfo {volume} {07}},\
  \bibinfo {pages} {120} (\bibinfo {year} {2018})},\ \Eprint
  {http://arxiv.org/abs/1805.02938} {arXiv:1805.02938 [hep-th]} \BibitemShut
  {NoStop}%
\bibitem [{\citenamefont {Chen}\ \emph
  {et~al.}(2021{\natexlab{b}})\citenamefont {Chen}, \citenamefont {Zhang},
  \citenamefont {Li}, \citenamefont {Hou},\ and\ \citenamefont
  {Huang}}]{Chen:2020ath}%
  \BibitemOpen
  \bibfield  {author} {\bibinfo {author} {\bibfnamefont {X.}~\bibnamefont
  {Chen}}, \bibinfo {author} {\bibfnamefont {L.}~\bibnamefont {Zhang}},
  \bibinfo {author} {\bibfnamefont {D.}~\bibnamefont {Li}}, \bibinfo {author}
  {\bibfnamefont {D.}~\bibnamefont {Hou}}, \ and\ \bibinfo {author}
  {\bibfnamefont {M.}~\bibnamefont {Huang}},\ }\href {\doibase
  10.1007/JHEP07(2021)132} {\bibfield  {journal} {\bibinfo  {journal} {JHEP}\
  }\textbf {\bibinfo {volume} {07}},\ \bibinfo {pages} {132} (\bibinfo {year}
  {2021}{\natexlab{b}})},\ \Eprint {http://arxiv.org/abs/2010.14478}
  {arXiv:2010.14478 [hep-ph]} \BibitemShut {NoStop}%
\bibitem [{\citenamefont {Zhou}\ \emph {et~al.}(2020)\citenamefont {Zhou},
  \citenamefont {Chen}, \citenamefont {Zhao},\ and\ \citenamefont
  {Ping}}]{Zhou:2020ssi}%
  \BibitemOpen
  \bibfield  {author} {\bibinfo {author} {\bibfnamefont {J.}~\bibnamefont
  {Zhou}}, \bibinfo {author} {\bibfnamefont {X.}~\bibnamefont {Chen}}, \bibinfo
  {author} {\bibfnamefont {Y.-Q.}\ \bibnamefont {Zhao}}, \ and\ \bibinfo
  {author} {\bibfnamefont {J.}~\bibnamefont {Ping}},\ }\href {\doibase
  10.1103/PhysRevD.102.086020} {\bibfield  {journal} {\bibinfo  {journal}
  {Phys. Rev. D}\ }\textbf {\bibinfo {volume} {102}},\ \bibinfo {pages}
  {086020} (\bibinfo {year} {2020})},\ \Eprint
  {http://arxiv.org/abs/2006.09062} {arXiv:2006.09062 [hep-ph]} \BibitemShut
  {NoStop}%
\bibitem [{\citenamefont {Hashimoto}\ \emph
  {et~al.}(2018{\natexlab{a}})\citenamefont {Hashimoto}, \citenamefont
  {Sugishita}, \citenamefont {Tanaka},\ and\ \citenamefont
  {Tomiya}}]{Hashimoto:2018ftp}%
  \BibitemOpen
  \bibfield  {author} {\bibinfo {author} {\bibfnamefont {K.}~\bibnamefont
  {Hashimoto}}, \bibinfo {author} {\bibfnamefont {S.}~\bibnamefont
  {Sugishita}}, \bibinfo {author} {\bibfnamefont {A.}~\bibnamefont {Tanaka}}, \
  and\ \bibinfo {author} {\bibfnamefont {A.}~\bibnamefont {Tomiya}},\ }\href
  {\doibase 10.1103/PhysRevD.98.046019} {\bibfield  {journal} {\bibinfo
  {journal} {Phys. Rev. D}\ }\textbf {\bibinfo {volume} {98}},\ \bibinfo
  {pages} {046019} (\bibinfo {year} {2018}{\natexlab{a}})},\ \Eprint
  {http://arxiv.org/abs/1802.08313} {arXiv:1802.08313 [hep-th]} \BibitemShut
  {NoStop}%
\bibitem [{\citenamefont {Akutagawa}\ \emph {et~al.}(2020)\citenamefont
  {Akutagawa}, \citenamefont {Hashimoto},\ and\ \citenamefont
  {Sumimoto}}]{Akutagawa:2020yeo}%
  \BibitemOpen
  \bibfield  {author} {\bibinfo {author} {\bibfnamefont {T.}~\bibnamefont
  {Akutagawa}}, \bibinfo {author} {\bibfnamefont {K.}~\bibnamefont
  {Hashimoto}}, \ and\ \bibinfo {author} {\bibfnamefont {T.}~\bibnamefont
  {Sumimoto}},\ }\href {\doibase 10.1103/PhysRevD.102.026020} {\bibfield
  {journal} {\bibinfo  {journal} {Phys. Rev. D}\ }\textbf {\bibinfo {volume}
  {102}},\ \bibinfo {pages} {026020} (\bibinfo {year} {2020})},\ \Eprint
  {http://arxiv.org/abs/2005.02636} {arXiv:2005.02636 [hep-th]} \BibitemShut
  {NoStop}%
\bibitem [{\citenamefont {Hashimoto}\ \emph
  {et~al.}(2018{\natexlab{b}})\citenamefont {Hashimoto}, \citenamefont
  {Sugishita}, \citenamefont {Tanaka},\ and\ \citenamefont
  {Tomiya}}]{Hashimoto:2018bnb}%
  \BibitemOpen
  \bibfield  {author} {\bibinfo {author} {\bibfnamefont {K.}~\bibnamefont
  {Hashimoto}}, \bibinfo {author} {\bibfnamefont {S.}~\bibnamefont
  {Sugishita}}, \bibinfo {author} {\bibfnamefont {A.}~\bibnamefont {Tanaka}}, \
  and\ \bibinfo {author} {\bibfnamefont {A.}~\bibnamefont {Tomiya}},\ }\href
  {\doibase 10.1103/PhysRevD.98.106014} {\bibfield  {journal} {\bibinfo
  {journal} {Phys. Rev. D}\ }\textbf {\bibinfo {volume} {98}},\ \bibinfo
  {pages} {106014} (\bibinfo {year} {2018}{\natexlab{b}})},\ \Eprint
  {http://arxiv.org/abs/1809.10536} {arXiv:1809.10536 [hep-th]} \BibitemShut
  {NoStop}%
\bibitem [{\citenamefont {Yan}\ \emph {et~al.}(2020)\citenamefont {Yan},
  \citenamefont {Wu}, \citenamefont {Ge},\ and\ \citenamefont
  {Tian}}]{Yan:2020wcd}%
  \BibitemOpen
  \bibfield  {author} {\bibinfo {author} {\bibfnamefont {Y.-K.}\ \bibnamefont
  {Yan}}, \bibinfo {author} {\bibfnamefont {S.-F.}\ \bibnamefont {Wu}},
  \bibinfo {author} {\bibfnamefont {X.-H.}\ \bibnamefont {Ge}}, \ and\ \bibinfo
  {author} {\bibfnamefont {Y.}~\bibnamefont {Tian}},\ }\href {\doibase
  10.1103/PhysRevD.102.101902} {\bibfield  {journal} {\bibinfo  {journal}
  {Phys. Rev. D}\ }\textbf {\bibinfo {volume} {102}},\ \bibinfo {pages}
  {101902} (\bibinfo {year} {2020})},\ \Eprint
  {http://arxiv.org/abs/2004.12112} {arXiv:2004.12112 [hep-th]} \BibitemShut
  {NoStop}%
\bibitem [{\citenamefont {Hashimoto}\ \emph {et~al.}(2022)\citenamefont
  {Hashimoto}, \citenamefont {Ohashi},\ and\ \citenamefont
  {Sumimoto}}]{Hashimoto:2021ihd}%
  \BibitemOpen
  \bibfield  {author} {\bibinfo {author} {\bibfnamefont {K.}~\bibnamefont
  {Hashimoto}}, \bibinfo {author} {\bibfnamefont {K.}~\bibnamefont {Ohashi}}, \
  and\ \bibinfo {author} {\bibfnamefont {T.}~\bibnamefont {Sumimoto}},\ }\href
  {\doibase 10.1103/PhysRevD.105.106008} {\bibfield  {journal} {\bibinfo
  {journal} {Phys. Rev. D}\ }\textbf {\bibinfo {volume} {105}},\ \bibinfo
  {pages} {106008} (\bibinfo {year} {2022})},\ \Eprint
  {http://arxiv.org/abs/2108.08091} {arXiv:2108.08091 [hep-th]} \BibitemShut
  {NoStop}%
\bibitem [{\citenamefont {Song}\ \emph {et~al.}(2021)\citenamefont {Song},
  \citenamefont {Oh}, \citenamefont {Ahn},\ and\ \citenamefont
  {Kima}}]{Song:2020agw}%
  \BibitemOpen
  \bibfield  {author} {\bibinfo {author} {\bibfnamefont {M.}~\bibnamefont
  {Song}}, \bibinfo {author} {\bibfnamefont {M.~S.~H.}\ \bibnamefont {Oh}},
  \bibinfo {author} {\bibfnamefont {Y.}~\bibnamefont {Ahn}}, \ and\ \bibinfo
  {author} {\bibfnamefont {K.-Y.}\ \bibnamefont {Kima}},\ }\href {\doibase
  10.1088/1674-1137/abfc36} {\bibfield  {journal} {\bibinfo  {journal} {Chin.
  Phys. C}\ }\textbf {\bibinfo {volume} {45}},\ \bibinfo {pages} {073111}
  (\bibinfo {year} {2021})},\ \Eprint {http://arxiv.org/abs/2011.13726}
  {arXiv:2011.13726 [physics.class-ph]} \BibitemShut {NoStop}%
\bibitem [{\citenamefont {Zhou}\ \emph {et~al.}(2023)\citenamefont {Zhou},
  \citenamefont {Wang}, \citenamefont {Pang},\ and\ \citenamefont
  {Shi}}]{Zhou:2023pti}%
  \BibitemOpen
  \bibfield  {author} {\bibinfo {author} {\bibfnamefont {K.}~\bibnamefont
  {Zhou}}, \bibinfo {author} {\bibfnamefont {L.}~\bibnamefont {Wang}}, \bibinfo
  {author} {\bibfnamefont {L.-G.}\ \bibnamefont {Pang}}, \ and\ \bibinfo
  {author} {\bibfnamefont {S.}~\bibnamefont {Shi}},\ }\href {\doibase
  10.1016/j.ppnp.2023.104084} {\bibfield  {journal} {\bibinfo  {journal} {Prog.
  Part. Nucl. Phys.}\ }\textbf {\bibinfo {volume} {104084}},\ \bibinfo {pages}
  {2023} (\bibinfo {year} {2023})},\ \Eprint {http://arxiv.org/abs/2303.15136}
  {arXiv:2303.15136 [hep-ph]} \BibitemShut {NoStop}%
\bibitem [{\citenamefont {Ahn}\ \emph {et~al.}(2024)\citenamefont {Ahn},
  \citenamefont {Jeong}, \citenamefont {Kim},\ and\ \citenamefont
  {Yun}}]{Ahn:2024gjf}%
  \BibitemOpen
  \bibfield  {author} {\bibinfo {author} {\bibfnamefont {B.}~\bibnamefont
  {Ahn}}, \bibinfo {author} {\bibfnamefont {H.-S.}\ \bibnamefont {Jeong}},
  \bibinfo {author} {\bibfnamefont {K.-Y.}\ \bibnamefont {Kim}}, \ and\
  \bibinfo {author} {\bibfnamefont {K.}~\bibnamefont {Yun}},\ }\href {\doibase
  10.1007/JHEP03(2024)141} {\bibfield  {journal} {\bibinfo  {journal} {JHEP}\
  }\textbf {\bibinfo {volume} {03}},\ \bibinfo {pages} {141} (\bibinfo {year}
  {2024})},\ \Eprint {http://arxiv.org/abs/2401.00939} {arXiv:2401.00939
  [hep-th]} \BibitemShut {NoStop}%
\bibitem [{\citenamefont {Gu}\ \emph {et~al.}(2024)\citenamefont {Gu},
  \citenamefont {Yan},\ and\ \citenamefont {Wu}}]{Gu:2024lrz}%
  \BibitemOpen
  \bibfield  {author} {\bibinfo {author} {\bibfnamefont {Z.-F.}\ \bibnamefont
  {Gu}}, \bibinfo {author} {\bibfnamefont {Y.-K.}\ \bibnamefont {Yan}}, \ and\
  \bibinfo {author} {\bibfnamefont {S.-F.}\ \bibnamefont {Wu}},\ }\href@noop {}
  {\  (\bibinfo {year} {2024})},\ \Eprint {http://arxiv.org/abs/2401.09946}
  {arXiv:2401.09946 [hep-th]} \BibitemShut {NoStop}%
\bibitem [{\citenamefont {Chen}\ and\ \citenamefont
  {Huang}(2024{\natexlab{a}})}]{Chen:2024ckb}%
  \BibitemOpen
  \bibfield  {author} {\bibinfo {author} {\bibfnamefont {X.}~\bibnamefont
  {Chen}}\ and\ \bibinfo {author} {\bibfnamefont {M.}~\bibnamefont {Huang}},\
  }\href {\doibase 10.1103/PhysRevD.109.L051902} {\bibfield  {journal}
  {\bibinfo  {journal} {Phys. Rev. D}\ }\textbf {\bibinfo {volume} {109}},\
  \bibinfo {pages} {L051902} (\bibinfo {year} {2024}{\natexlab{a}})},\ \Eprint
  {http://arxiv.org/abs/2401.06417} {arXiv:2401.06417 [hep-ph]} \BibitemShut
  {NoStop}%
\bibitem [{\citenamefont {Chen}\ and\ \citenamefont
  {Huang}(2024{\natexlab{b}})}]{Chen:2024mmd}%
  \BibitemOpen
  \bibfield  {author} {\bibinfo {author} {\bibfnamefont {X.}~\bibnamefont
  {Chen}}\ and\ \bibinfo {author} {\bibfnamefont {M.}~\bibnamefont {Huang}},\
  }\href@noop {} {\  (\bibinfo {year} {2024}{\natexlab{b}})},\ \Eprint
  {http://arxiv.org/abs/2405.06179} {arXiv:2405.06179 [hep-ph]} \BibitemShut
  {NoStop}%
\bibitem [{\citenamefont {Zhu}\ \emph {et~al.}(2025)\citenamefont {Zhu},
  \citenamefont {Chen}, \citenamefont {Zhou}, \citenamefont {Zhang},\ and\
  \citenamefont {Huang}}]{Zhu:2025gxo}%
  \BibitemOpen
  \bibfield  {author} {\bibinfo {author} {\bibfnamefont {L.}~\bibnamefont
  {Zhu}}, \bibinfo {author} {\bibfnamefont {X.}~\bibnamefont {Chen}}, \bibinfo
  {author} {\bibfnamefont {K.}~\bibnamefont {Zhou}}, \bibinfo {author}
  {\bibfnamefont {H.}~\bibnamefont {Zhang}}, \ and\ \bibinfo {author}
  {\bibfnamefont {M.}~\bibnamefont {Huang}},\ }\href@noop {} {\  (\bibinfo
  {year} {2025})},\ \Eprint {http://arxiv.org/abs/2501.17763} {arXiv:2501.17763
  [hep-ph]} \BibitemShut {NoStop}%
\bibitem [{\citenamefont {Critelli}\ \emph {et~al.}(2017)\citenamefont
  {Critelli}, \citenamefont {Noronha}, \citenamefont {Noronha-Hostler},
  \citenamefont {Portillo}, \citenamefont {Ratti},\ and\ \citenamefont
  {Rougemont}}]{Critelli:2017oub}%
  \BibitemOpen
  \bibfield  {author} {\bibinfo {author} {\bibfnamefont {R.}~\bibnamefont
  {Critelli}}, \bibinfo {author} {\bibfnamefont {J.}~\bibnamefont {Noronha}},
  \bibinfo {author} {\bibfnamefont {J.}~\bibnamefont {Noronha-Hostler}},
  \bibinfo {author} {\bibfnamefont {I.}~\bibnamefont {Portillo}}, \bibinfo
  {author} {\bibfnamefont {C.}~\bibnamefont {Ratti}}, \ and\ \bibinfo {author}
  {\bibfnamefont {R.}~\bibnamefont {Rougemont}},\ }\href {\doibase
  10.1103/PhysRevD.96.096026} {\bibfield  {journal} {\bibinfo  {journal} {Phys.
  Rev. D}\ }\textbf {\bibinfo {volume} {96}},\ \bibinfo {pages} {096026}
  (\bibinfo {year} {2017})},\ \Eprint {http://arxiv.org/abs/1706.00455}
  {arXiv:1706.00455 [nucl-th]} \BibitemShut {NoStop}%
\bibitem [{\citenamefont {Zhang}\ and\ \citenamefont
  {Huang}(2022)}]{Zhang:2022uin}%
  \BibitemOpen
  \bibfield  {author} {\bibinfo {author} {\bibfnamefont {L.}~\bibnamefont
  {Zhang}}\ and\ \bibinfo {author} {\bibfnamefont {M.}~\bibnamefont {Huang}},\
  }\href {\doibase 10.1103/PhysRevD.106.096028} {\bibfield  {journal} {\bibinfo
   {journal} {Phys. Rev. D}\ }\textbf {\bibinfo {volume} {106}},\ \bibinfo
  {pages} {096028} (\bibinfo {year} {2022})},\ \Eprint
  {http://arxiv.org/abs/2209.00766} {arXiv:2209.00766 [nucl-th]} \BibitemShut
  {NoStop}%
\bibitem [{\citenamefont {Matsumoto}\ and\ \citenamefont
  {Nakamura}(2018)}]{Matsumoto:2018ukk}%
  \BibitemOpen
  \bibfield  {author} {\bibinfo {author} {\bibfnamefont {M.}~\bibnamefont
  {Matsumoto}}\ and\ \bibinfo {author} {\bibfnamefont {S.}~\bibnamefont
  {Nakamura}},\ }\href {\doibase 10.1103/PhysRevD.98.106027} {\bibfield
  {journal} {\bibinfo  {journal} {Phys. Rev. D}\ }\textbf {\bibinfo {volume}
  {98}},\ \bibinfo {pages} {106027} (\bibinfo {year} {2018})},\ \Eprint
  {http://arxiv.org/abs/1804.10124} {arXiv:1804.10124 [hep-th]} \BibitemShut
  {NoStop}%
\bibitem [{\citenamefont {Wang}\ \emph {et~al.}(2019)\citenamefont {Wang},
  \citenamefont {Wei}, \citenamefont {Li},\ and\ \citenamefont
  {Huang}}]{Wang:2018sur}%
  \BibitemOpen
  \bibfield  {author} {\bibinfo {author} {\bibfnamefont {X.}~\bibnamefont
  {Wang}}, \bibinfo {author} {\bibfnamefont {M.}~\bibnamefont {Wei}}, \bibinfo
  {author} {\bibfnamefont {Z.}~\bibnamefont {Li}}, \ and\ \bibinfo {author}
  {\bibfnamefont {M.}~\bibnamefont {Huang}},\ }\href {\doibase
  10.1103/PhysRevD.99.016018} {\bibfield  {journal} {\bibinfo  {journal} {Phys.
  Rev. D}\ }\textbf {\bibinfo {volume} {99}},\ \bibinfo {pages} {016018}
  (\bibinfo {year} {2019})},\ \Eprint {http://arxiv.org/abs/1808.01931}
  {arXiv:1808.01931 [hep-ph]} \BibitemShut {NoStop}%
\bibitem [{\citenamefont {Chen}\ \emph
  {et~al.}(2021{\natexlab{c}})\citenamefont {Chen}, \citenamefont {Huang},\
  and\ \citenamefont {Liao}}]{Chen:2021aiq}%
  \BibitemOpen
  \bibfield  {author} {\bibinfo {author} {\bibfnamefont {H.-L.}\ \bibnamefont
  {Chen}}, \bibinfo {author} {\bibfnamefont {X.-G.}\ \bibnamefont {Huang}}, \
  and\ \bibinfo {author} {\bibfnamefont {J.}~\bibnamefont {Liao}},\ }\href
  {\doibase 10.1007/978-3-030-71427-7_11} {\bibfield  {journal} {\bibinfo
  {journal} {Lect. Notes Phys.}\ }\textbf {\bibinfo {volume} {987}},\ \bibinfo
  {pages} {349} (\bibinfo {year} {2021}{\natexlab{c}})},\ \Eprint
  {http://arxiv.org/abs/2108.00586} {arXiv:2108.00586 [hep-ph]} \BibitemShut
  {NoStop}%
\bibitem [{\citenamefont {Fujimoto}\ \emph {et~al.}(2021)\citenamefont
  {Fujimoto}, \citenamefont {Fukushima},\ and\ \citenamefont
  {Hidaka}}]{Fujimoto:2021xix}%
  \BibitemOpen
  \bibfield  {author} {\bibinfo {author} {\bibfnamefont {Y.}~\bibnamefont
  {Fujimoto}}, \bibinfo {author} {\bibfnamefont {K.}~\bibnamefont {Fukushima}},
  \ and\ \bibinfo {author} {\bibfnamefont {Y.}~\bibnamefont {Hidaka}},\ }\href
  {\doibase 10.1016/j.physletb.2021.136184} {\bibfield  {journal} {\bibinfo
  {journal} {Phys. Lett. B}\ }\textbf {\bibinfo {volume} {816}},\ \bibinfo
  {pages} {136184} (\bibinfo {year} {2021})},\ \Eprint
  {http://arxiv.org/abs/2101.09173} {arXiv:2101.09173 [hep-ph]} \BibitemShut
  {NoStop}%
\bibitem [{\citenamefont {Nunes}\ \emph {et~al.}(2024)\citenamefont {Nunes},
  \citenamefont {Farias}, \citenamefont {Tavares},\ and\ \citenamefont
  {Tim\'oteo}}]{Nunes:2024hzy}%
  \BibitemOpen
  \bibfield  {author} {\bibinfo {author} {\bibfnamefont {R.~M.}\ \bibnamefont
  {Nunes}}, \bibinfo {author} {\bibfnamefont {R.~L.~S.}\ \bibnamefont
  {Farias}}, \bibinfo {author} {\bibfnamefont {W.~R.}\ \bibnamefont {Tavares}},
  \ and\ \bibinfo {author} {\bibfnamefont {V.~S.}\ \bibnamefont {Tim\'oteo}},\
  }\href@noop {} {\  (\bibinfo {year} {2024})},\ \Eprint
  {http://arxiv.org/abs/2412.14541} {arXiv:2412.14541 [hep-ph]} \BibitemShut
  {NoStop}%
\bibitem [{\citenamefont {Chen}\ \emph
  {et~al.}(2024{\natexlab{b}})\citenamefont {Chen}, \citenamefont {Fu},
  \citenamefont {Huang},\ and\ \citenamefont {Ma}}]{Chen:2024utf}%
  \BibitemOpen
  \bibfield  {author} {\bibinfo {author} {\bibfnamefont {H.-L.}\ \bibnamefont
  {Chen}}, \bibinfo {author} {\bibfnamefont {W.-j.}\ \bibnamefont {Fu}},
  \bibinfo {author} {\bibfnamefont {X.-G.}\ \bibnamefont {Huang}}, \ and\
  \bibinfo {author} {\bibfnamefont {G.-L.}\ \bibnamefont {Ma}},\ }\href@noop {}
  {\  (\bibinfo {year} {2024}{\natexlab{b}})},\ \Eprint
  {http://arxiv.org/abs/2410.20704} {arXiv:2410.20704 [hep-ph]} \BibitemShut
  {NoStop}%
\bibitem [{\citenamefont {Chen}\ \emph
  {et~al.}(2024{\natexlab{c}})\citenamefont {Chen}, \citenamefont {Wang},
  \citenamefont {Hou},\ and\ \citenamefont {Ren}}]{Chen:2024edy}%
  \BibitemOpen
  \bibfield  {author} {\bibinfo {author} {\bibfnamefont {J.-X.}\ \bibnamefont
  {Chen}}, \bibinfo {author} {\bibfnamefont {S.}~\bibnamefont {Wang}}, \bibinfo
  {author} {\bibfnamefont {D.-F.}\ \bibnamefont {Hou}}, \ and\ \bibinfo
  {author} {\bibfnamefont {H.-C.}\ \bibnamefont {Ren}},\ }\href@noop {} {\
  (\bibinfo {year} {2024}{\natexlab{c}})},\ \Eprint
  {http://arxiv.org/abs/2410.04763} {arXiv:2410.04763 [hep-ph]} \BibitemShut
  {NoStop}%
\end{thebibliography}%
\end{document}